\documentclass[12pt]{article}

\begin{document}

\title{Quantization as Asymptotics of Diffusion Processes in the Phase Space}
\author{E. M. Beniaminov}
\date{}
\maketitle
\begin{abstract}
This work is an extended version of the paper~\cite{ben}, in which the main results were
announced.
We consider certain classical diffusion process for a wave function on the phase space.
It is shown that at the time of order $10^{-11}$ {\it sec} this process converges to a process
considered by quantum mechanics and described by the Schrodinger equation.
This model studies the probability distributions in the phase space corresponding to the
wave functions of quantum mechanics.
We estimate the parameters of the model using the Lamb--Retherford experimental data
on shift in the spectrum of hydrogen atom and the assumption on the heat reason of the
considered diffusion process.

In the paper it is shown that the quantum mechanical description of the processes
can arise as an approximate description of more exact models.
For the model considered in this paper, this approximation arises
when the Hamilton function changes slowly under deviations of coordinates, momenta, and time
on intervals whose length is of order determined by the Planck constant
and by the diffusion intensities.
\end{abstract}

\tableofcontents

\section{Introduction}
In this paper we propose a model which, on the one hand, allows one to estimate
the probability distribution of a quantum particle in the phase space in the low temperature
heat field. For the first time this problem was solved by Wigner \cite{wigner}, but
he constructed ``quasi-distributions'' on the phase space which can be negative and hence
have no physical sense. On the other hand, the proposed model
yields one more construction of quantization of mechanical systems and can be used in
the new approach to foundation of the classical quantization procedure.
This is an old problem. Various approaches to this problem, in particular probabilistic ones,
can be found in
\cite{bohm_vigier, nelson, pena_cetto, baublitz, maslov1, maslov2}. These works essentially
influenced the author during the construction of the present model.

In this paper we consider the classical model of a diffusion process for a wave
(complex valued)
function on the phase space.   The analysis of the differential equation of the model
shows that the motion in the model splits into rapid and slow motions.
The result of the rapid motion is that the system, starting from an arbitrary
wave function on the phase space, goes to a function belonging to certain distinguished
subspace. The elements of this subspace are parameterized by the wave functions
depending only on the coordinates. The slow motion along the subspace is described by the
Schrodinger equation.

Using the assumptions on the heat reason of the diffusions and the correspondence of
the consequences of the model with the known physical experiments of
Lamb--Retherford \cite{lamb}
(the Lamb shift in the spectrum of hydrogen atom), we estimate the diffusion
coefficients and the time of the transition process from the classical description,
in which the Heisenberg indeterminacy principle does not hold, to the quantum description
in which the Heisenberg principle already holds. The time of the transition process
has order $1/T \cdot 10^{-11} sec$, where
$T$ is the temperature of the medium.

The results of this work have been announced in the paper \cite{ben}. Proofs of
theorems~4 and~5 are instructive but rather technical, hence they are exposed in the Appendix.
The estimate of the parameters of the model is also exposed in the Appendix.

The author is grateful to professor G.~L.~Litvinov, who was attentive to this work,
made a lot of editorial comments, and stimulated an essential revision of the text,
and to professor A.~V.~Stoyanovsky, who translated this paper to English.

\section{Description of the model}
We consider a mathematical model of a process whose state at each moment of time
is given by a wave function, which is a complex valued function
$\varphi (x, p)$, where $(x,p) \in R^{2n},$
and $n$  is the dimension of configuration space.
In contrast to quantum mechanics, where the wave function depends only on coordinates or
only on momenta, in our case the wave function depends both on coordinates and on
momenta.
As in quantum mechanics, it is assumed that for the wave functions the superposition
principle holds, and the probability density
$\rho (x,p)$ on the phase space, corresponding to the wave function
$\varphi(x, p),$ is given by the standard formula
\begin{equation}\label{rho}
\rho (x,p)= \varphi^*(x, p) \varphi(x, p)=|\varphi(x, p)|^2.
\end{equation}

In the present work we consider the classical model of diffusion process for the wave function
$\varphi (x, p)$ on the phase space. It is assumed that each complex vector of the wave
function is simultaneously in 4 motions:

the base point of the complex vector moves along the classical trajectory given by the
Hamilton function $H(x, p)$;

the base point of the vector moves randomly with respect to coordinates and momenta, being in
diffusion process with constant diffusion coefficients $a^2$ and $b^2$
with respect to coordinates and momenta, respectively;

the base point of each vector moves along a random trajectory as a result of motions described
in the two preceding points, and the vector itself rotates with very large constant
angular velocity $\omega ={mc^2}/{\hbar}$ in the coordinate system related with this point,
where $m$ is the mass of the particle, $c$ is the light velocity, ${\hbar}$ is the Planck
constant;

the length of all complex vectors of the wave function at the moment $t$ of time is
multiplied by $\exp (abnt/\hbar)$ (this is a purely technical requirement which does
not affect the relative probabilities of position of the particle in the phase space).

It is assumed that the wave vector $\varphi (x, p, t)$
at the point $(x, p)$ at the moment $t$ of time equals, by the superposition principle,
to the sum of wave vectors given by the distribution of vectors $\varphi (x, p, 0)$ at the
initial moment of time which get to the point $(x,p)$ at the moment $t$ due to
the motions described above.

\section{The main results}
\subsection{The mathematical model of the process}
Consider the diffusion process on the phase space in which the wave function
$\varphi (x, p, t)$ at the moment $t$  satisfies the differential equation
\begin{equation}\label{eq_diff}
\frac{\partial\varphi}{\partial{t}}=\sum_{k=1}^{n}
\biggl(
\frac{\partial H}{\partial x_k} \frac{\partial\varphi}{\partial p_k}-
\frac{\partial H}{\partial p_k} \frac{\partial\varphi}{\partial x_k}
\biggr)
-\frac{i}{\hbar}
\biggl(H-\sum_{k=1}^{n}\frac{\partial H}{\partial p_k}p_k\biggr)\varphi
+\Delta_{a,b}{\varphi},
\end{equation}
\begin{equation}\label{delta}
\mbox{where }\ \ \ \ \ \Delta_{a,b}{\varphi}=
a^2\sum_{k=1}^{n}\biggl(\frac{\partial}{\partial{x_k}}-
     \frac{ip_k}{\hbar}\biggr)^{2}\varphi
+b^2\sum_{k=1}^{n}\frac{\partial^2 }{\partial{p^2_k}}\varphi
     +\frac {abn}{\hbar}{\varphi},
\end{equation}
where
$H(x, p)$ is the Hamilton function;
$a^2$ and $b^2$ are the diffusion coefficients with respect to coordinates and momenta,
respectively.

If we omit the last summand in equation~(\ref{eq_diff}), then we obtain a first order
partial differential equation
${\partial\varphi}/{\partial{t}}= A  \varphi$, where
\begin{equation}\label{eq_det}
 A \varphi=\sum_{k=1}^{n}
\biggl(
\frac{\partial H}{\partial x_k} \frac{\partial\varphi}{\partial p_k}-
\frac{\partial H}{\partial p_k} \frac{\partial\varphi}{\partial x_k}
\biggr)
-\frac{i}{\hbar}
\biggl(H-\sum_{k=1}^{n}\frac{\partial H}{\partial p_k}p_k\biggr)\varphi.
\end{equation}

This part of equation~(\ref{eq_diff}) describes the deterministic component of the motion of
complex vectors $\varphi(x, p,t)$ along the characteristics of the equation.
According to the equation, in this motion the base point
of each vector moves along the classical trajectory given by the Hamiltonian $H(x, p)$,
and the vector itself rotates at each point of the trajectory with the angular
velocity
$\omega'= \frac{1}{\hbar}\biggl(H-\sum_{k=1}^{n}\frac{\partial H}{\partial p_k}p_k\biggr).$

Note that in the case when the configuration space is three dimensional and
$H=E= c\sqrt {m^2c^2 +p^2}$,  we have
$$\omega' dt  = \frac{1}{\hbar}\biggl(H-\sum_{k=1}^{n}\frac{\partial H}{\partial p_k}p_k\biggr) dt =
\frac{mc^2}{\hbar}\frac{mc^2dt}{H}= \frac{mc^2}{\hbar} d\tau ,$$
where $\tau=mc^2dt/H$,  in accordance with the formulas of special relativity theory, is the
proper time in the coordinate system related with the particle moving
with the momentum $p$.  I.~e. in this case, the vector whose base point moves
along the classical trajectory, rotates with the constant angular velocity
$\omega ={mc^2}/{\hbar}$ in the coordinate system related with this point.

On the contrary, if in the right hand side of equation~(\ref{eq_diff}) we leave only
the last summand
of the form~(\ref{delta}), then we obtain the equation
\begin{equation}\label{eq_delta}
\frac{\partial\varphi}{\partial{t}}=
a^2\sum_{k=1}^{n}\biggl(\frac{\partial}{\partial{x_k}}-
     \frac{ip_k}{\hbar}\biggr)^{2}\varphi
+b^2\sum_{k=1}^{n}\frac{\partial^2 }{\partial{p^2_k}}\varphi
     +\frac {abn}{\hbar}{\varphi}.
\end{equation}
This equation describes the diffusion component of the motion of
vectors $\varphi(x, p,t)$ in the phase space. In this motion, the base points of the vectors
move according to the classical homogeneous diffusion process with the diffusion coefficients
with respect to coordinates and momenta equal to $a^2$ and $b^2$,
respectively. And the vector itself is parallel transported during small random transports from
a point $(x, p)$
to the point $(x+dx, p+dp)$, and its length at moment $t$ is multiplied by $\exp (abnt/\hbar)$.
Note that the parallel transport of vectors on the phase space is given by a connection
expressed by the following formula:
$
L_{(dx,dp)}\varphi(x, p) - \varphi(x, p)\approx  -({i}/{\hbar})\varphi(x, p) p dx,
$
where $L_{(dx,dp)}\varphi(x, p)$ is the parallel transport of the vector $\varphi(x, p)$
from the point $(x, p)$ along the infinitely small vector $(dx, dp)$.

In the particular case when the configuration space is three dimensional, such the connection on
the phase space is related to synchronization of moving clocks at the points of the
phase space. Indeed, if a particle with coordinates $x=(x_1;  x_2;  x_3)$  moves with the
velocity $v=(v_1;   v_2;   v_3)$,   then, according to formulas of special relativity theory,
proper time is expressed through the observer time $t$ by the formula
\begin{eqnarray}\label{tauv}
\tau=\frac{t-  x v /{c^2}}{\sqrt{1-{v^2}/{c^2}}} ,
\end{eqnarray}
where $\   x v =x_1v_1+x_2v_2+x_3v_3$ is the scalar product of the vectors $x$ and $v$, and
$c$ is the velocity of light.

For a free particle with the momentum $p=(p_1; p_2; p_3)$ and the stationary mass $m$,
the energy $ E=c\sqrt{p^2+m^2c^2}$ and, respectively,
\begin{equation}\label{v}
v = \frac{pc}{\sqrt{p^2+m^2c^2}},\ \ \ \ \
\sqrt{1-\frac{v^2}{c^2}}  =  \frac{mc}{\sqrt{p^2+m^2c^2}}.
\end{equation}

   Substituting these expressions into (\ref{tauv}),  after computations we obtain:
\begin{equation}\label{tau}
 \tau=\frac{Et- x p}{mc^2}.
\end{equation}

Consider the distribution of complex vectors $\varphi(x,p,t)$ on the phase space,
rotating with constant angular velocity $\omega ={mc^2}/{\hbar}$ in the proper time $\tau$,
which, at the moment $\tau=0$, are equal to one and the same vector $\varphi_0$. We have
\begin{equation}
 \varphi(x,p,t)=\varphi_0 \exp\biggl(-i\frac{mc^2}{\hbar}\tau\biggr).
\end{equation}
Substituting formula (\ref{tau}) into this expression,  we obtain
\begin{equation}
 \varphi(x,p,t)=\varphi_0 \exp\biggl(\frac{-i(Et- x p)}{\hbar}\biggr).
\end{equation}
Hence, if $L_{(\triangle x,0)}\varphi(x,p,t)$ is the shift of the vector
$\varphi(x,p,t)=\varphi_0 \exp(-imc^2 \tau/{\hbar})$ by the vector $\triangle x$
along coordinates $x$ without change of proper time, then
 \begin{equation}
 L_{(\triangle x,0)}\varphi(x,p,t)=\varphi(x,p,t) \exp\biggl(\frac{-i \triangle x p}{\hbar}\biggr).
\end{equation}
In the limit of infinitely small $\triangle x$ we obtain the required formula for this case:
$
L_{(dx,0)}\varphi(x, p) - \varphi(x, p) =  -({i}/{\hbar})\varphi(x, p) p dx,
$
where $dx$ is the infinitely small shift of coordinates $x$.

On the other hand, if we have a shift $L_{(0,\triangle p)}$ of the vector
$\varphi(x,p,t)=\varphi_0 \exp(-imc^2 \tau/{\hbar})$
along momentum by $\triangle p$ without change of proper time, then, since in the
special relativity approximation, acceleration does not change the proper time of a particle,
we have the equality
$ L_{(0,\triangle p)}\varphi(x,p,t)=\varphi(x,p,t)$ and
$L_{(0,dp)}\varphi(x, p) - \varphi(x, p) = 0.$

Hence, by linearity of $L_{(dx,dp)}$ with respect to $(dx,dp)$, we obtain the required equality
in the general case:
$
L_{(dx,dp)}\varphi(x, p) - \varphi(x, p) =  -({i}/{\hbar})\varphi(x, p) p dx.
$

 Note that under these assumptions, the derivation with respect to the vector of infinitely
 small shift along the $k$-th coordinate corresponds to the differential operator
 $ D_{\displaystyle x_k}= {\partial}/{\partial{x_k}}-{ip_k}/{\hbar}$,
and the derivation with respect to the shift along the $k$-th momentum corresponds to the usual
differential operator
$ D_{\displaystyle p_k}= {\partial}/{\partial{p_k}}, $ where $ k=1,...,n. $

Note also that these operators of shift along coordinates and momenta do not
commute. The commutators of these differential operators read
    $$ \left[D_{\displaystyle p_k},D_{\displaystyle x_k}\right]=
       -{i}/{\hbar}\mbox{  and  }
     \left[D_{\displaystyle p_k},D_{\displaystyle x_j}\right]=0,
       \mbox{ where } k\neq j \mbox{ and }  k,j=1,...,n.$$
Thus, the shifts along coordinates and momenta of wave functions on the phase space realize
a representation of the Heisenberg group.

\subsection{Analysis of the diffusion component of the equation}
Consider the diffusion equation~(\ref{eq_delta}) in more detail.
This equation can be represented as follows:

\begin{equation}\label{diff1}
 \frac{\partial\varphi}{\partial{t}}=\Delta_{a,b}{\varphi}=
a^2\Delta_x{\varphi}+
b^2\Delta_p{\varphi}+\frac{nab}{\hbar}{\varphi},
\end{equation}
where
$$
\Delta_x{\varphi}=\sum_{k=1}^{n}D_{\displaystyle x_k}^2{\varphi}
  =\sum_{k=1}^{n}\biggl(\frac{\partial^2\varphi}{\partial{x^2_k}}-
     \frac{2ip_k}{\hbar}\frac{\partial\varphi}{\partial{x_k}}-
     \frac{p_{k}^{2}}{\hbar^2}\varphi \biggl),
$$
$$
\Delta_p{\varphi}=\sum_{k=1}^{n}D_{\displaystyle p_k}^2{\varphi}
= \sum_{k=1}^n\frac{\partial^2\varphi}{\partial{p_k^2}}.
$$

It is natural to call the operator $\Delta_{a,b}$ by the diffusion operator for
the representation of the Heisenberg group with the diffusion intensities $a$ and $b$
with respect to coordinates and momenta, respectively.

Let us look for a solution of equation (\ref{diff1}) in the form
\begin{equation}\label{view1}
 \varphi(x,p,t)=
\varphi^{0}(x,p,t)\
\exp\biggl(\frac{i x p }{\hbar}\biggr).
\end{equation}

 Substituting this expression into equation (\ref{diff1}),  dividing both parts of the
 equality by $\exp({i x p}/{\hbar}) $  and transferring the summand
$({nab}/{\hbar}){\varphi^0}$ into the left hand side,
we obtain the equation
\begin{equation} \label{EQ1}
     \frac{\partial\varphi^0}{\partial{t}}-\frac{nab}{\hbar}{\varphi^0}
=
\sum_{k=1}^{n}\Biggl(a^2\frac{\partial^2\varphi^0}{\partial{x^2_k}}+
b^2\biggl(\frac{\partial^2\varphi^0}{\partial{p^2_k}}+
     \frac{2ix_k}{\hbar}\frac{\partial\varphi^0}{\partial{p_k}}-
     \frac{x_{k}^{2}}{\hbar^2}\varphi^0\biggr)\Biggr),
\end{equation}
     where $\varphi^0=\varphi^0(x,p,t)$ is some function.

     To solve equation~(\ref{EQ1}), let us decompose the function $\varphi^0(x,p,t) $
into the Fourier integral with respect to $p$, i.~e., let us represent the function
$\varphi^0(x,p,t) $ in the form
\begin{equation} \label{fur}
  \varphi^0(x,p,t) ={\mathcal F}_{\hbar}\psi^{0}(x,y,t)=\frac{1}{(2\pi{\hbar})^{n/2}}
\int_{R^n}\psi^{0}(x,y,t)\exp\biggl(-{\frac{i y p}{\hbar}}\biggr)dy,
\end{equation}
\begin{equation} \label{fur_1}
\mbox{where   }\ \
  \psi^0(x,y,t) ={\mathcal F}_{\hbar}^{-1}\varphi^0(x,p,t)=\frac{1}{(2\pi{\hbar})^{n/2}}
\int_{R^n}\varphi^{0}(x,p,t)\exp\biggl({\frac{i y p }{\hbar}}\biggr)dp.
\end{equation}
     Substituting this expression for $\varphi^0(x,p,t)$   into the equation (\ref{EQ1}),
we obtain that $\psi^0(x,y,t)$ satisfies the equation
\begin{equation} \label{EQpsi0}
     \frac{\partial\psi^0}{\partial{t}}-\frac{nab}{\hbar}{\psi^0}
=
\sum_{k=1}^{n}\biggl(a^2\frac{\partial^2\psi^0}{\partial{x^2_k}}-
\frac{b^2(x_k-y_k)^2}{\hbar^2}\psi^0\biggr)
\end{equation}
The right hand side of this equation is a self adjoint operator with discrete
spectrum consisting of negative numbers.

Indeed, the equation for eigenvalues of this operator reads
\begin{equation} \label{gar}
\sum_{k=1}^{n}\biggl(a^2\frac{\partial^2\chi}{\partial{x^2_k}}-
\frac{b^2(x_k-y_k)^2}{\hbar^2}\chi\biggr)
=\lambda\cdot\chi,
\end{equation}
where $\chi=\chi(x,y)$ is a function of $x$ and $y$.
This equation is the stationary Schrodinger equation for harmonic oscillator,
and it is well studied (see, for instance, \cite{landau3}, p.~94).

In particular, it is known that on the set of functions which tend to zero as $x$
tends to infinity, equation~(\ref{gar}) has
discrete spectrum consisting of negative eigenvalues
$\lambda_1>\lambda_2\geq\ldots$.    The greatest eigenvalue
$\lambda_1=-{nab}/{\hbar}$   corresponds to the eigenfunction
$\chi_1(x,y)=
({b}/{a\pi\hbar})^{n/4}\exp\left(-{b}(x-y)^2/{2a\hbar}   \right)$.
The next eigenvalues are less than $\lambda_1$, and the difference is greater than or equal
to $ab/\hbar$.

Since the eigenfunctions $\chi_k(x,y)$ of the operator~(\ref{gar})
form a complete system of functions in the class of functions tending to zero as
$x$ tends to infinity,  an arbitrary function
$\psi^0(x,y,t) $ from this class can be represented as a series
 $$
\psi^0(x,y,t)=\sum_{k=1}^{\infty}c_k(y,t)\chi_k(x,y),
 $$
\begin{equation}\label{c_k}
\mbox{where}\ \  c_k(y,t)=\int_{R^n} \psi^0(x,y,t)\chi_k(x,y) dx
\end{equation}
are the coefficients of the decomposition of the function $\psi^0(x,y,t)$
with respect to eigenfunctions $\chi_k(x,y)$.

Substituting the expression of the function $\psi^0(x,y,t) $ in the form of this
series into equation~(\ref{EQpsi0}),   we obtain that this equation in the orthonormal basis
of eigenfunctions $\chi_k(x,y),$  $k=1, 2, \ldots$, splits into an infinite system
of equations:
  $$
     \frac{\partial c_k(y,t)}{\partial{t}}=
\left(\lambda_k+\frac{nab}{\hbar}\right) c_k(y,t)
     \qquad k=1,2,\ldots
  $$
where $\lambda_1+{nab}/{\hbar}=0$ and
$\lambda_k+{nab}/{\hbar}\leq -{ab}/{\hbar}$ for $k>1.$

    Hence $c_1(y,t)=c_1(y,0)$, and
$c_k(y,t)=c_k(y,0)\exp((\lambda_k+{nab}/{\hbar})t)$
exponentially decay with time for $k>1$. Hence the summand in $\psi^0$
corresponding to the first eigenvalue will give the main contribution into the function
$\psi^0$ after time of order $(\hbar/ab)$.

Thus, we have obtained that a solution of equation~(\ref{EQpsi0}),
after time $t$ of order
$\hbar/ab$, becomes exponentially close to the function $\psi^0(x,y)$, where
\begin{eqnarray}\label{psi_0}
\psi^0(x,y)= \lim_{t \to \infty}\psi^0(x,y,t) =
c_1(y,0)\chi_1(x,y).
\end{eqnarray}
Respectively, since by definition
$\varphi^0(x,p,t) ={\mathcal F}_{\hbar}\psi^{0}(x,y,t)$ and Fourier transform is continuous,
we have
\begin{eqnarray}\label{def_varphi_0}
\varphi^0(x,p)\stackrel {def}{=}
{\mathcal F}_{\hbar}\psi^0(x,y)= \lim_{t \to \infty}{\mathcal F}_{\hbar}\psi^0(x,y,t) =
{\mathcal F}_{\hbar}(c_1(y,0)\chi_1(x,y)).
\end{eqnarray}

Since we will not use other eigenfunctions, introduce the notation
\begin{eqnarray}
\label{def_chi}
\chi(x-y)\stackrel {def}{=}\chi_1(x,y)=
({b}/{a\pi\hbar})^{n/4}\exp\left(-{b}(x-y)^2/{2a\hbar} \right).
\end{eqnarray}

To make the formulas shorter, let us also denote
$\psi(y)\stackrel{def}{=}c_1(y,0)$, where $c_1(y,0)$ is expressed by the formula~(\ref{c_k})
with $k=1$ and $t=0$.
That is,
\begin{eqnarray}\label{def_psi}
\psi(y)=\int_{R^n} \psi^0(x,y,0)\chi_1(x,y) dx =\int_{R^n} \psi^0(x,y,0)\chi(x-y) dx.
\end{eqnarray}
Since by formula~(\ref{view1}) $\varphi(x,p,t)=\varphi^0(x,p,t)\exp(ixp/\hbar),$
then by formulas (\ref{def_varphi_0}),  (\ref{def_chi}) and using the notation
$\psi(y)\stackrel{def}{=}c_1(y,0)$ above and also the equality~(\ref{def_psi})
and notation (\ref{fur_1}), we obtain the following theorem.

{\it {\bf Theorem~1}.
Let $\varphi(x,p,0)$ be an arbitrary function such that Fourier transform of the
function $\varphi(x,p,0)\exp(-ixp/\hbar)$  with respect to $p$ tends to zero as
$x \rightarrow  \infty $.
Then the solution $\varphi(x,p,t)$ of the diffusion equation~(\ref{eq_delta})
exponentially with time (with the number in the exponent equal to $-abt/{\hbar}$) tends to
a stationary solution of the form
\begin{equation}\label{view_varphi}
\varphi_0(x,p)=\lim_{t \to \infty} \varphi(x,p,t)=
\frac{1}{(2\pi{\hbar})^{n/2}}
\!\int\limits_{R^n}\!\!\psi(y)\chi(x-y)
e^{-{{i  (y-x)p}/{\hbar}}}
dy,
\end{equation}
\begin{eqnarray}\label{psi}
 \mbox {where }\ \ \psi(y) =\frac{1}{(2\pi{\hbar})^{n/2}}
\int\limits_{R^{2n}}\!\!\varphi(x,p,0)
e^{{{i  (y-x)p}/{\hbar}}}
\chi(x-y)dpdx,
\end{eqnarray}
\begin{eqnarray}\label{chi}
 \mbox{and          }\ \ \ \ \ \ \ \ \ \ \ \ \ \ \ \ \
  \chi(x-y)=\left(\frac{b}{a\pi\hbar}\right)^{n/4}e^{-{{b}(x-y)^2}/{(2a\hbar)}}.
\end{eqnarray}}

Note that $\chi^2(x-y)$   is the probability density of the normal distribution
with respect to $x$  with the mathematical expectation $y$ and dispersion $a\hbar/(2b)$.
If the quantity
$a\hbar/(2b)$ is small, then the function $\chi^2(x-y)$ is close to the delta function of
$x-y$.

The composition of expressions~(\ref{psi}) and (\ref{view_varphi}) yields
a projector $P_0$ form the space of wave functions defined on the phase space
onto certain subspace.
The elements of this subspace are parameterized by functions of the form $\psi(y),$
where $y\in R^n$,
i.~e., by wave functions on the configuration space.

{\it {\bf Theorem~2}.
The projection operator $P_0$ given by composition of expressions~(\ref{psi}) and
(\ref{view_varphi}),
has the form
\begin{eqnarray}\label{p_0}
P_0\varphi =
\frac{1}{(2\pi{\hbar})^{n}}
\!\int\limits_{R^{2n}}\!\!\
\varphi(x',p')e^{-\frac{b(x'-x)^2}{4a\hbar}}e^{-\frac{a(p'-p)^2}{4b\hbar}}
e^{\frac{i  (x-x')(p+p')}{2\hbar}} dx'dp'.
\end{eqnarray}
The operator $P_0$ is self-adjoint and commutes with the operator $\Delta_{a,b}$.
}

Proof of this theorem is obtained by substitution of formulas~(\ref{psi}) and (\ref{chi})
into (\ref{view_varphi}),
by an algebraic transformation of the number in the exponent,
and by computation of the integral over $y$. The integral over $y$
is the Fourier transform of the exponent of a quadratic polynomial,
whose analytical expression is known. The author performed the computations using the system
Mathematica \cite{math}, supporting symbolic mathematical computations.
The commutativity of the operators $P_0$ and $\Delta_{a,b}$ follows from the fact that
the orthogonal projector $P_0$ distinguishes the subspace of eigenvectors
of the self-adjoint operator $\Delta_{a,b}$ with the zero eigenvalue.

Formulas~(\ref{view_varphi}) and (\ref{rho}) imply the following statement.

{\it {\bf Theorem~3}.
If $\psi(x)$ is a wave function on the configuration space and $\varphi(x,p)$
is the wave function on the phase space corresponding to it by formula~(\ref{view_varphi}),
then the probability density in the phase space $\rho(x,p)=|\varphi(x,p)|^2$ is expressed
by the formula
\begin{eqnarray}\label{rho_psi}
\rho (x,p)=
\frac{1}{(2\pi{\hbar})^{n}} \left(\frac{b}{4a\pi\hbar}\right)^{n/2}
\int\limits_{R^{2n}} \! \psi\!
\left(x+\frac{x''-x'}{2}\right)\psi^{*}\!\left(x+\frac{x''+x'}{2}\right)  \\ \nonumber
\exp\!\left({-\frac{b(x'')^2}{4a\hbar}}\right)
\exp\!\left({-\frac{b(x')^2}{4a\hbar}}\right)
\exp\!\left({\frac{i x' p}{\hbar}}\right)
dx'' dx'.
\end{eqnarray}
}

In contrast to quasi-distributions
$$W(x,p)=\frac{1}{(2\pi{\hbar})^{n}}
\int\limits_{R^{n}}\!\!\psi\!\left(x-\frac{x'}{2}\right)\psi^{*}\!\left(x+\frac{x'}{2}\right)
\exp\!\left({\frac{i x' p}{\hbar}}\right)dx'
$$
defined by Wigner~\cite{wigner},
the density $\rho(x,p)$ in the phase space, given by expression~(\ref{rho_psi}), is
always nonnegative.
Its expression differs from the expression for the Wigner function by exponents under
integral, which give the densities of distributions close to delta functions.

To prove Theorem~3, one should substitute into formula~(\ref{rho})
expression~(\ref{view_varphi}). We obtain
\begin{equation}\label{rho_psi_chi}
\rho (x,p)=\varphi\varphi^*=
\frac{1}{(2\pi{\hbar})^{n}}
\int\limits_{R^{2n}}\!\!\psi(y)\psi^{*}(y')\chi(x-y)\chi(x-y')
e^{-{{i (y-y')p}/{\hbar}}}
dy dy',
\end{equation}
where $\chi(x-y)$ is given by relation~(\ref{chi}). After substitution of (\ref{chi})
into (\ref{rho_psi_chi}),
the change of coordinates
$y=x+(x''-x')/2$ and $y'=x+(x''+x')/2$ under integral and a transformation of the numbers
in the exponents, we obtain
formula (\ref{rho_psi}).

The algebra of observables given by real functions on the
phase space but averaged over the probability densities of the form~(\ref{rho_psi}),
has been studied in~\cite{ben_arxiv}.

The function $\rho(x)$ of density of probability distribution in the configuration space
is expressed through the density $\rho(x,p)$  in the phase space by the formula
$\rho(x)=\int_{R^n}\rho(x,p) dp.$
Hence, integrating expression~(\ref{rho_psi_chi}) over $p$,
we obtain the following statement.

{\it  {\bf Corollary 1}.
If $\psi(x)$ is a wave function on the configuration space, then
the corresponding probability density
 $\rho(x)$   in the configuration space is given by the formula
 \begin{equation}\label{rho_x}
\rho (x)=\int_{R^n}|\psi(y)|^2 \chi^2(x-y) dy,
\end{equation}
where $\chi(x-y)$ is given by~(\ref{chi}).
That is, $\rho (x)$ is obtained from $|\psi(x)|^2$ by smoothing (convolution)
with the density of normal distribution with dispersion $a\hbar/(2b)$,
and the exactness of defining coordinate is bounded by the quantity $\sim \sqrt{a\hbar/(2b)}$.
}

As it is known  \cite{landau4},
in quantum electrodynamics the minimal error of measuring coordinates of an electron in the
stationary system is bounded by the quantity $\hbar/(mc),$ where $m$ is the mass of the
electron, and $c$ is the light velocity.
Hence the statement of Corollary~1, although not corresponding to non-relativistic
quantum mechanics (in which it is assumed that coordinates can be measured with any degree
of exactness), does not contradict with a more exact theory, quantum electrodynamics.

If one assumes that the diffusion is induced by heat action on the electron, then
the diffusion coefficients with respect to coordinates and momenta are expressed in
statistical physics (see, for example, \cite{isihara}, Ch.7, \S 4 and \S 9)
through the temperature $T$ by the formulas
 $a^2= kT/(m\gamma) \ \  \mbox{      and     }\ \ b^2=\gamma kTm,$
where $k$ is the Boltzmann constant,  $m$ is the mass of the electron, $\gamma$ is the
friction coefficient of the medium per unit of mass.
Hence $a/b=(\gamma m)^{-1}$ and $ab=kT$.
That is, in this case, the quantity $a/b$, which enters expression~(\ref{chi}) and
determines the dispersion of smoothing in Corollary~1, does not depend on the
temperature.
On the other hand, the time $t$ of the transformation process determined in Theorem~1,
has the form $t \sim {\hbar}/{(ab)}={\hbar}/{(kT)}= T^{-1}\cdot 7.638\cdot 10^{-12}\ ñ.$
More detailed formulas for the estimate of the quantity $\hbar/(ab)$ are given in
Appendix~3.

\subsection{Analysis of the model of the process}
Let us return to the study of the main equation~(\ref{eq_diff}). Taking into account the
estimate made at the end of the previous subsection,
let us consider the quantity ${\hbar}/{(ab)}$ in equation~(\ref{eq_diff})
as a small parameter, and let us assume that coordinates and momenta change a little
at this time in the classical motion defined by the Hamiltonian
$H(x, p)$, and also let us assume that the function $H(x,p)$
and all its derivatives grow at infinity no faster than a polynomial.

{\it {\bf Theorem 4}.
The motion described by equation~(\ref{eq_diff}) asymptotically splits as
${\hbar}/{(ab)} \rightarrow   0$ into a rapid motion and a slow one.

1) As a result of rapid motion, an arbitrary wave function $\varphi(x, p, 0)$
turns at the time of order ${\hbar}/{(ab)}$ into a function of the form~(\ref{view_varphi}):
\begin{eqnarray}\label{view_varphi'}
\varphi(x, p)=\frac{1}{(2\pi{\hbar})^{n/2}}
\!\int\limits_{R^n}\!\!\psi(y)\chi(x-y)
e^{-{{i  (y-x)p}/{\hbar}}}
dy,\\
\label{chi_def'}\mbox{where          }\ \ \ \ \ \ \ \ \ \
  \chi(x-y)=\left(\frac{b}{a\pi\hbar}\right)^{n/4}e^{-{{b}(x-y)^2}/{(2a\hbar)}}.
\end{eqnarray}
 The wave functions of the form~(\ref{view_varphi'}) form a linear subspace.
Elements of this subspace are parameterized by wave functions $\psi(y)$ depending only on
coordinates $y\in R^n$.

2) The slow motion starting from a nonzero wave function $\varphi(x, p, 0)$ of the form
~(\ref{view_varphi'}) from this subspace, goes inside this subspace, and is parameterized by
the wave function $\psi(y,t)$ depending on time. The function $\psi(y,t)$ satisfies the
Schrodinger equation of the form
$ i\hbar {\partial \psi}/{\partial t} = \hat{H}\psi$,  where
\begin{eqnarray}\label{hat_H}
\hat{H}\psi & = & \frac {1}{(2\pi \hbar)^n} \int \limits_{R^{3n}}
\biggl (H(x, p)-\sum_{k=1}^{n}\biggl(\frac{ \partial H}{\partial x_k}
+\frac {i b}{a}\frac{ \partial H}{\partial p_k}\biggr)(x_k-y'_k)
\biggr) \\
   &  & \times \chi(x- y) \chi(x- y')
 e^{\frac{i}{\hbar}(y-y') p}  \psi(y',t) dy' dx dp, \nonumber
\end{eqnarray}
and $\chi(x- y) $ is given by formula~(\ref{chi_def'}).
\it}

Proof of Part 1 of Theorem~4 is postponed till Appendix, due to its large volume and
technicalities.

{\bf  Proof of Part 2 of Theorem~4.}  In the first Part of Theorem~4 it is stated that
after the rapid motion, the initial distribution $\varphi(x,p,0)$ turns to the form
with just a small difference with~(\ref{view_varphi'}).  After the slow motion
the distribution remains in the class of functions of the form~(\ref{view_varphi'}),
but changes in time.

     To study the slow motion, let us look for a solution of equation~(\ref{eq_diff})
in the form~(\ref{view_varphi'}) in which $\psi=\psi(y',t)$
is considered as dependent on time.

Let us substitute expression~(\ref{view_varphi'}) into equation~(\ref{eq_diff}). Since
by construction, this expression for $\varphi$ satisfies equation
$\Delta_{a,b}\varphi=0$, then after substitution of expression~(\ref{view_varphi'}) into
equation~(\ref{eq_diff}) and after dividing both parts of the equation by
$ \exp\left({i x p }/{\hbar} \right)$, we obtain the following equation:
\begin{eqnarray}\label{slow1}
\lefteqn{\frac{1}{(2\pi{\hbar})^{n/2}}
 \int \limits_{R^n}
\frac{\partial\psi(y',t)}{\partial t}\chi(x-y')
\exp\biggl(-{\frac{i y'  p }{\hbar}}\biggr)dy'}
\nonumber\\
& &=\frac{1}{(2\pi{\hbar})^{n/2}}
\int \limits_{R^n}
\biggl[ \sum_{k=1}^n \biggl (\frac{\partial H}{\partial x_k}\biggl(-\frac{i(y'_k-x_k)}{\hbar}\biggr)
-\frac{\partial H}{\partial p_k}
\biggl(-\frac{b}{a\hbar}(x_k-y'_k)
+\frac{i}{\hbar}p_k\biggr)\biggr)
\nonumber
\\
& &-\frac{i}{\hbar}\biggl(H
-\sum_{k=1}^n\frac{\partial H}{\partial p_k}p_k\biggr)
\biggr]
\psi(y',t)\chi(x-y')
\exp\biggl(-{\frac{i y'p }{\hbar}}\biggr)dy'.
\end{eqnarray}
If in the obtained equation one opens the brackets, reduces the similar terms, multiplies
both parts of the equation by
$1/(2\pi\hbar)^{n/2}\chi(x-y) \exp(i y p /\hbar)$
and integrates by $p$ and by $x$, then, taking into account the equality
$\int_{R^n} \chi^{2}(x-y)dx=1$,
we obtain the following equation:
\begin{equation} \label{fineq}
\frac{\partial\psi}{\partial t}=-\frac{i}{\hbar}\hat H \psi
\ \ \ \mbox{    or   }\ \ \
i\hbar\frac{\partial\psi}{\partial t}=\hat H \psi,
\end{equation}
where the operator $ \hat H$  is obtained from the function $H$ by formula~(\ref{hat_H}),
required in Part 2 of Theorem~4.

{\it{\bf Theorem~5}. If $\frac {a\hbar}{b}$ is a small quantity and
$H(x, p) = \frac{p^2}{2m}+ V(x),$
then the operator $\hat H$, up to terms of order $a\hbar/b$, has the form
\begin{equation}\label{hatH}
\hat H \approx - \frac{\hbar^2}{2m}\biggl(\sum_{k=1}^{n}\frac{\partial^2 }{\partial{y^2_k}}\biggr)
+V(y)-\frac{a\hbar}{4b}\sum_{k=1}^{n}\frac{\partial^2 V}{\partial{y^2_k}} +\frac{3nb\hbar}{4ma}.
\end{equation}}

Proof of Theorem~5 is given in Appendix~2.

The first two summands in formula~(\ref{hatH}) give the standard Hamilton operator.
 The last summand is a constant and can be neglected. The previous summand before the last one
 will be considered (due to the smallness of $a\hbar/b$)
as a perturbation of the Hamilton operator.

Assuming that the deviations in the spectrum of the hydrogen atom (the Lamb shift)
observed in the Lamb--Retherford experiments~\cite{lamb} are induced by the previous to the
last summand in formula~(\ref{hatH}),
one can estimate the quantity $a/b$. The computations by the standard method of
perturbation theory analogous to the computations of \cite{welt}, give the following
estimate: $a/b =3.41\cdot 10^4 sec/g$ (see the computations in Appendix~3).
 Hence, the standard deviation of the normal distribution $\chi^2$, with which we make
 smoothing in formula~(\ref{rho_x}), has the form
$\sqrt{a\hbar/(2b)}=4.24 \cdot 10^{-12} cm.$
This quantity is much less than the radius of the hydrogen atom, and it is close to the
Compton wave length of the electron $\hbar/(mc)=3.86\cdot 10^{-11}cm$.

\section{Conclusion}
In this paper it is shown that the standard quantum mechanical description of a process
can arise as an approximation of certain classical model for a diffusion process
for a wave function in the phase space.
The computations show that the proposed model in the form of differential
equation~(\ref{eq_diff}) describes physical processes in a rather adequate manner
in the standard cases for standard Hamiltonian. But this model can be applied as well
for computations of processes with a nonstandard Hamiltonian or a Hamiltonian
rapidly changing in time, as in sudden perturbations~\cite{dyihne}
or for periodically changing potential with frequency of order $ab/\hbar$,
and it can be compared with experimental data.

\newpage
\section*{APPENDICES}

\begin{flushright}
{\large Appendix 1}
\end{flushright}

\section*{Proof of Part 1 of Theorem 4}
\addcontentsline{toc}{section}{Appendix 1. Proof of Theorem 4}

Let $\varphi(x,p,t)$ be a solution of equation~(\ref{eq_diff}).
In the notations~(\ref{delta}) and (\ref{eq_det}) equation~(\ref{eq_diff}) takes the form
\begin{equation}\label{eq_short}
\frac{\partial\varphi}{\partial{t}}=A \varphi + \Delta_{a,b}{\varphi},
\end{equation}
where $A$ is a skew Hermitian operator, and $\Delta_{a,b}$ is the self-adjoint
diffusion operator.

Consider the derivative with respect to time of $||\varphi||^2$ (the square of the norm
of the function $\varphi$). We have
$$\frac{d ||\varphi||^2}{dt}=\frac{d \langle\varphi;\varphi \rangle}{dt}=
\langle\frac{\partial \varphi}{\partial t}; \varphi \rangle +
\langle\varphi ;\frac{\partial \varphi}{\partial t} \rangle=
\langle A \varphi + \Delta_{a,b}{\varphi}; \varphi \rangle
+ \langle\varphi ; A \varphi + \Delta_{a,b}{\varphi} \rangle.
$$
Hence, by linearity of the scalar product $ \langle \ ;\  \rangle$ and by the equalities
$ \langle A \varphi;\varphi \rangle=-\langle\varphi;A \varphi \rangle$ and
$ \langle\Delta_{a,b}\varphi;\varphi \rangle=\langle\varphi; \Delta_{a,b} \varphi \rangle$,
we obtain the equality
\begin{equation}\label{d_norm}
2||\varphi||\frac{d ||\varphi||}{dt}=\frac{d ||\varphi||^2}{dt}=2 \langle
\Delta_{a,b}\varphi;\varphi \rangle
\ \ \mbox{ or }
\ \ \frac{d ||\varphi||}{dt}=\frac{1}{ ||\varphi||} \langle \Delta_{a,b}\varphi;\varphi \rangle.
\end{equation}

Denote by ${\bar \varphi }\stackrel{def}{=}\varphi/||\varphi||$ the normalized function $\varphi$.
If the function $\varphi$ satisfies equation~(\ref{eq_short}), then, taking into account
formula~(\ref{d_norm}), we have
\begin{equation}\label{eq_norm}
\frac{\partial\bar \varphi}{\partial{t}}
=\frac{\partial}{\partial{t}}\left( \frac{\varphi}{||\varphi||}\right)=
\frac {1}{||\varphi||}\frac{\partial\varphi}{\partial{t}}-
\frac {\varphi}{||\varphi||^2}\frac{d||\varphi||}{d{t}}
=A \bar \varphi + \Delta_{a,b}{\bar \varphi}-
\langle \Delta_{a,b}\bar \varphi;\bar \varphi \rangle \bar \varphi .
\end{equation}

By definition of the function $\bar \varphi$ its norm $||\bar \varphi||\equiv 1$,
and it is natural to call equation~(\ref{eq_norm}) the equation for the normalized wave
function.

Consider now the projection of the function $\bar \varphi(x,p,t)$ on the subspace
of stationary solutions of the diffusion equation~(\ref{eq_delta}).

Let us represent the function $\bar \varphi(x,p,t)$ in the form
$\bar\varphi=\varphi_0+\varphi_1$, where
$\varphi_0=P_0 \bar{\varphi}$,
$\varphi_1=({\bf 1}-P_0)\bar{\varphi}=P_1 \bar{\varphi}$, and $P_0$ is the projector from
Theorem~2.
The self-adjointness of the projection operator $P_0$ implies the equalities
$\langle \varphi_0; \varphi_1 \rangle=0$ and
$||\varphi_0 ||^2+|| \varphi_1 ||^2=||\bar \varphi ||^2 =1.$

For the proof of the first Part of Theorem~4 one needs to show that the quantity
$|| \varphi_1 ||^2=1-||\varphi_0 ||^2$
becomes small after time of order $\hbar/(ab)$.

To make formulas shorter, introduce the notation
\begin{equation}\label{def_eta}
\eta (t)\stackrel {def}{=}||\varphi_0(t)||^2, \ \ \mbox{where}\ \ \varphi_0=P_0 \bar\varphi.
\end{equation}

{\bf Statement 1.} {\it If $\varphi$ satisfies equation~(\ref{eq_short}), then the quantity
$\eta (t)=||\varphi_0 ||^2$
satisfies the differential equation
\begin{equation}\label{eq_eta}
\dot \eta = \alpha (t) \sqrt{1-\eta }\sqrt{\eta } +\beta(t)(1-\eta ){\eta },
\end{equation}
where $\beta(t)\geq \beta_{min}=2ab /\hbar $ and
$|\alpha (t)|\leq \alpha_{max}
=2\cdot  \max_{\bar \varphi_0}|| A \bar \varphi_0-P_0 A \bar \varphi_0||$.
The maximum in the latter expression is taken over all normalized functions
$\bar \varphi_0$ from the subspace of stationary solutions of the diffusion
equation~(\ref{eq_delta}), for which the function $|\bar \varphi_0(x,p)|^2$
gives a probability distribution in the physical region of the phase space for the
given process.
}

To prove Statement 1, let us consider how the quantity $||\varphi_0||^2$ changes in time,
where $\varphi_0=P_0 \bar \varphi$  and $\varphi$ satisfies equation~(\ref{eq_short}).

We have the following equalities which follow from the definition of the function $\varphi_0$,
from independence of the operator $P_0$  of time,
from equality~(\ref{eq_norm}), from linearity of the scalar product
$\langle \ ;\ \rangle$ with respect to each argument,
from commutativity of $P_0$ with $\Delta_{a,b}$,
from self-adjointness of the operators $P_0$ and $\Delta_{a,b}$, from skew Hermitian
property of the operator $A$ and from the projection property $P_0=P_0^2$:
\begin{eqnarray}\label{eq_eta_1}
\dot \eta& =&\frac {d \langle \varphi_0; \varphi_0 \rangle}{d t}
=\frac {d \langle P_0 \bar \varphi; P_0 \bar \varphi \rangle}{d t}=
  \langle P_0 \frac{\partial \bar \varphi}{\partial t}; P_0 \bar \varphi \rangle +
 \langle P_0\bar \varphi; P_0 \frac{\partial \bar \varphi}{\partial t} \rangle\nonumber\\
&=& \langle  \frac{\partial \bar \varphi}{\partial t}; P_0^2 \bar \varphi \rangle +
\langle  P_0^2 \bar \varphi;  \frac{\partial \bar \varphi}{\partial t} \rangle=
\langle \frac{\partial \bar\varphi}{\partial t}; \varphi_0 \rangle +
 \langle  \varphi_0; \frac{\partial \bar\varphi}{\partial t} \rangle\nonumber\\
&=&\Bigl\langle A \bar \varphi + \Delta_{a,b}{\bar \varphi}-  \langle \Delta_{a,b}\bar \varphi;
\bar \varphi \rangle \bar \varphi;\  \varphi_0 \Bigr\rangle +
 \Bigl\langle  \varphi_0;\  A \bar \varphi + \Delta_{a,b}{\bar \varphi}-
 \langle \Delta_{a,b}\bar \varphi;\bar \varphi \rangle \bar \varphi \Bigr\rangle\nonumber\\
&&\mbox{(substitute expression for } \bar\varphi=\varphi_0+\varphi_1, \mbox{ use }\nonumber\\
&&\mbox{the linearity of operators and equalities } A^*=-A,\
\Delta_{a,b}{\varphi_0}=0 )\nonumber\\
&=&- \langle \varphi_1; A \varphi_0 \rangle - \langle A \varphi_0;  \varphi_1 \rangle
-2 \langle \Delta_{a,b}\varphi_1; \varphi_1 \rangle\cdot  \langle \varphi_0;
\varphi_0 \rangle\nonumber\\
&=&-\left \langle \frac{\varphi_1}{||\varphi_1||};
A\frac{\varphi_0}{||\varphi_0||}\right \rangle \cdot ||\varphi_1||\cdot ||\varphi_0|| -
\left \langle A \frac{\varphi_0}{||\varphi_0||}; \frac{\varphi_1}{||\varphi_1||}
\right \rangle\cdot ||\varphi_1||\cdot ||\varphi_0||\nonumber\\
&&-2 \left \langle \Delta_{a,b}\frac{\varphi_1}{||\varphi_1||};
\frac{\varphi_1}{||\varphi_1||} \right \rangle\cdot  ||\varphi_1||^2\cdot ||\varphi_0||^2
\nonumber\\
&=&-\! \left(\langle \bar \varphi_1; A\bar \varphi_0 \rangle
\!+ \!\langle A \bar \varphi_0; \bar \varphi_1 \rangle\right)\!||\varphi_1||\  ||\varphi_0||
\!-\!2 \langle \Delta_{a,b}\bar \varphi_1;
\bar \varphi_1 \rangle  ||\varphi_1||^2 ||\varphi_0||^2,
\end{eqnarray}
where $\bar \varphi_1$ and $\bar \varphi_0$ are the normalized functions $\varphi_1$ and
$\varphi_0$, respectively.

Since $\eta =||\varphi_0||^2$ and $||\varphi_0 ||^2+|| \varphi_1 ||^2=||\bar \varphi ||^2 =1$,
then $|| \varphi_1 ||=\sqrt{1-\eta }$.

Substituting these equalities into equation~(\ref{eq_eta_1}) and introducing notations
$$
\alpha (t)\stackrel{def}{=}-\left(\langle \bar \varphi_1; A\bar \varphi_0 \rangle +
\langle A \bar \varphi_0; \bar \varphi_1 \rangle\right)
\mbox{ and }
\beta (t)\stackrel{def}{=}-2 \langle \Delta_{a,b}\bar \varphi_1; \bar \varphi_1 \rangle,
$$
we obtain the equation from Statement 1:
$\dot \eta=\alpha (t) \sqrt{1-\eta }\sqrt{\eta } +\beta(t)(1-\eta ){\eta }.$

For the estimate of $|\alpha (t)|$, note that since
$\langle \bar \varphi_1;
A\bar \varphi_0 \rangle = \langle A \bar \varphi_0; \bar \varphi_1 \rangle^*$, where
at the end of the equality the star *
denotes complex conjugation, then $\alpha (t)= 2\cdot  Re (\langle -A \bar \varphi_0;
\bar \varphi_1 \rangle)$, where $Re$ denotes the real part of a complex number.
Hence we obtain the following estimates:
\begin{eqnarray}
|\alpha (t)|= 2  |Re (\langle -A \bar \varphi_0; \bar \varphi_1 \rangle)|\leq2|\langle A \bar \varphi_0; \bar \varphi_1 \rangle|
=2|\langle  A \bar \varphi_0-P_0 A \bar \varphi_0; \bar \varphi_1 \rangle |=&&\nonumber\\
2|\langle ( A - P_0 A )\bar \varphi_0; \bar \varphi_1 \rangle |\leq 2||( A -P_0 A )
\bar \varphi_0||\ ||\bar \varphi_1 ||
=2||A\bar \varphi_0 - P_0 A \bar \varphi_0||.&&
\end{eqnarray}

The latter inequality in the previous formula is obtained using the Cauchy--Schwartz
inequality estimating the absolute value of the scalar product.
The function $-P_0 A \bar \varphi_0$, which is orthogonal to the function $\varphi_1$,
has been added to $A \bar \varphi_0$, in order to obtain the projection of the function
$A \bar \varphi_0$ onto the subspace orthogonal to the space of stationary solutions
of the diffusion equation~(\ref{eq_delta}).

Let us now estimate the quantity $\beta (t){=}-2 \langle \Delta_{a,b}\bar \varphi_1;
\bar \varphi_1 \rangle$. To this end, decompose the function
$\bar \varphi_1$ with respect to the orthonormal eigenfunctions $\chi_i $ of the self-adjoint
operator $\Delta_{a,b}$. We have
$\bar \varphi_1= \sum_{i=0}^\infty  c_i \chi_i$. Since the vector $\bar \varphi_1$
is orthogonal to the vectors with eigenvalue equal to 0,
we have $ \Delta_{a,b}\bar \varphi_1=\sum_{i:\ \lambda _i \not=0}\lambda_i   c_i \chi_i$
and $\sum_{i:\ \lambda _i \not=0}   c_i^2=1. $ Hence
\begin{eqnarray}
\langle \Delta_{a,b}\bar \varphi_1; \bar \varphi_1 \rangle
=\langle \sum_{i:\ \lambda _i \not=0}\!\!\lambda_i   c_i \chi_i;
\sum_{i=1}^\infty  c_i \chi_i \rangle=
\!\sum_{i:\ \lambda _i \not=0}\!\! \lambda_i c_i^2
\leq \!\sum_{i:\ \lambda _i \not=0}\!\! \lambda_{max}  c_i^2=\lambda_{max},
\end{eqnarray}
where $\lambda_{max}$ is the maximal nonzero eigenvalue of the
self-adjoint operator $ \Delta_{a,b}$.

Using this inequality and also knowing that $\lambda_{max}=-ab/\hbar$ (see Proof of
Theorem 2), we obtain the estimate required in Statement~1:
$$
\beta (t){=}-2 \langle \Delta_{a,b}\bar \varphi_1; \bar \varphi_1 \rangle \geq  \beta_{min}=-2\lambda_{max}=\frac {2ab}{\hbar}.
$$
Statement~1 is proved.

{\bf Statement 2.} {\it As $\hbar\rightarrow 0$ the estimate of the quantity $\alpha_{max}$
has the form $\alpha_{max}=(C+o_{\hbar}(1))/\sqrt{\hbar}$,
where $o_{\hbar}(1)$ is an infinitely small quantity with respect to $\hbar$, and the
constant $C$ is determined by the maximal value of the absolute values of the first and second
derivatives of the Hamilton function $H(x, p)$, when the coordinates and momenta belong to
the physical region of values for the given process.}

Proof of Statement~2 will be postponed until the end of this section.

For the estimate of the time of the transformation process described in Theorem~4,
let us rewrite the equation from Statement~1 in the following form:
$$
dt= \frac{d\eta }{\alpha (t) \sqrt{1-\eta }\sqrt{\eta } +\beta(t)(1-\eta ){\eta }}\ \ \
\mbox{       or}
$$
\begin{equation}\label{int_t}
t=\!\!\int\limits_{\varepsilon}^{1-\varepsilon }
\!\!\frac{d\eta }{\alpha (t) \sqrt{1-\eta }\sqrt{\eta } +\beta(t)(1-\eta ){\eta }}\leq
\!\!\int\limits_{\varepsilon}^{1-\varepsilon }
\!\!\frac{d\eta }{-\alpha_{max} \sqrt{1-\eta }\sqrt{\eta } +\beta_{min}(1-\eta ){\eta }},
\end{equation}
where the integral is taken over the interval $(\varepsilon , 1-\varepsilon)$, on which
the denominator of the expression under the latter integral is positive.
Simple computations show that for this the following inequalities should hold:
\begin{equation}\label{varepsilon_bigger}
\frac{1}{2}> {\varepsilon}>\frac{1}{2}-\sqrt{\frac{1}{4}-\frac{\alpha_{max}^2}{\beta_{min}^2}}.
\end{equation}

If one takes into account that $\beta_{min}=2ab/\hbar $ (see Statement~1) and
$\alpha_{max}=(C+o_{\hbar}(1))/\sqrt{\hbar}$ (see Statement~2),
then, as $\hbar\rightarrow 0$, the right hand side of the previous inequality
can be represented in the following form by decomposing into the Taylor series:
$$
\frac{1}{2}-\sqrt{\frac{1}{4}-\frac{\alpha_{max}^2}{\beta_{min}^2}}=
\frac{\alpha_{max}^2}{\beta_{min}^2}+O\biggl(\frac{\alpha_{max}}{\beta_{min}}\biggr)^4=
\frac{C^2\hbar}{4a^2b^2}+o\biggl(\frac{\hbar}{ab}\biggr).
$$
This and inequality~(\ref{varepsilon_bigger}) imply that the number $\varepsilon$
can be chosen to be arbitrary (small, as $\hbar\rightarrow 0$), satisfying the
inequality
\begin{equation}\label{varepsilon_bigger_h}
\frac{1}{2}> {\varepsilon}>\frac{C^2\hbar}{4a^2b^2}+o\biggl(\frac{\hbar}{ab}\biggr).
\end{equation}

The last integral in inequality~(\ref{int_t}) of the form
\begin{equation}\label{t_varepsilon_def}
t_{\varepsilon}\stackrel{def}{=}
\int\limits_{\varepsilon}^{1-\varepsilon }\!\!\frac{d\eta }{-\alpha_{max} \sqrt{1-\eta }\sqrt{\eta } +\beta_{min}(1-\eta ){\eta }},
\end{equation}
has been computed using the system Mathematica 5.0 (see \cite{math}),
and the result was decomposed into the series with respect to $\beta_{min}$ as
$\beta_{min} \rightarrow \infty$. These computations yield the following equality:
\begin{equation}
t_{\varepsilon}=\frac{4\, arctanh(1-2\varepsilon)}{\beta_{min}}+
\frac{4(1-2\varepsilon )}{\sqrt{\varepsilon (1-\varepsilon )}}
\frac{\alpha_{max}}{\beta_{min}^2}+
O\left(\frac{1}{\beta^3_{min}}\right),
\end{equation}
where $ arctanh(x)=(ln(1+x)-ln(1-x))/2$ is the hyperbolic arctangens.

If one assumes that $\varepsilon $ and $\hbar$ are small quantities, then, substituting
into this equality the expressions for
$\beta_{min}=2ab/\hbar $ and $\alpha_{max}=(C+o_{\hbar}(1))/\sqrt{\hbar}$ from Statements~1
and 2, and decomposing the obtained expression into a series with respect to
$\varepsilon$, one can obtain the following estimate:
\begin{equation}\label{t_min}
t_{\varepsilon}=(-\ln\ \varepsilon +o(\varepsilon ) )\frac{\hbar}{ab}+
\frac{(1+o(\varepsilon ))}{\sqrt{\varepsilon}}\frac{C}{\sqrt{\hbar}}\biggl(\frac{\hbar}{ab}\biggr)^2+
O\biggl(\frac{\hbar^2}{a^2 b^2}\biggr).
\end{equation}

Inequality~(\ref{int_t}), formulas~(\ref{varepsilon_bigger_h}),  (\ref{t_varepsilon_def}),
(\ref{t_min}), and the definition of the function $\eta(t)$ (\ref{def_eta})
immediately imply the following statement.

{\bf Statement 3.}  {\it Let $\bar\varphi(x,p,t)$ be a normalized solution
of equation~(\ref{eq_diff}), and, at the initial moment for $t=0$, let the following
inequality hold:
$\eta(0)\stackrel{def}{=} ||P_0 \bar\varphi(x,p,0) ||^2\geq  \varepsilon$,
where $P_0$ is the projection operator from Theorem~2 and $\varepsilon $ is an
arbitrary number satisfying the inequalities
$$
\frac{1}{2}> {\varepsilon}>\frac{C^2\hbar}{4a^2b^2}+o\biggl(\frac{\hbar}{ab}\biggr).
$$
Then, for any $t>t_\varepsilon $,  where
\begin{equation}\label{t_min_h}
t_{\varepsilon}=(-\ln\ \varepsilon +o(\varepsilon ) )\frac{\hbar}{ab}+
\frac{(1+o(\varepsilon ))}{\sqrt{\varepsilon}}\frac{C}{\sqrt{\hbar}}\biggl(\frac{\hbar}{ab}
\biggr)^2+
O\biggl(\frac{\hbar^2}{a^2 b^2}\biggr),\nonumber
\end{equation}
the quantity $\eta (t)\stackrel{def}{=}||P_0\bar\varphi (x, p, t)||^2\geq 1-\varepsilon $,
i.~e., the square of the distance from the function
$\bar\varphi(x,p,t)$ to the subspace of stationary solutions of the diffusion equation,
described by Theorem~1,  will be less than or equal to $\varepsilon$.
}

Part 1) of Theorem~4, being proved at this section, immediately follows from
Statement~3. For conclusion of the proof it remains to prove Statement~2.

{\bf Proof of Statement~2.}
In the proof of Statement~2 we shall need computations of some integrals containing the
function $\chi(y)$ given by expression~(\ref{chi}) from Theorem~1. The results of computations
of these integrals are listed in the following two Lemmas.

{\bf Lemma 1.} {\it Let $\tilde \chi$ be the Fourier transform of the function $\chi$, where
$  \chi(y)=\left({b}/{a\pi\hbar}\right)^{n/4}exp{(-{{b}y^2}/{(2a\hbar)})}.$
Then the following equality holds:
\begin{equation}\label{lemma1_1}
\tilde\chi (k)=\frac{1}{(2\pi \hbar)^{n/2}}\int_{R^{n}}\chi (y)e^{\frac{iyk}{\hbar}}dy=
\left(\frac{a}{b\pi\hbar}\right)^{n/4}e^{-\frac{{a}k^2}{2b\hbar}}.
\end{equation}
On the contrary, the Fourier transform of $\tilde \chi$ yields $\chi$. I.~e.,
\begin{equation}\label{lemma1_2}
\chi (y)=\frac{1}{(2\pi \hbar)^{n/2}}\int_{R^{n}}\tilde\chi (k)e^{\frac{iyk}{\hbar}}dy.
\end{equation}
One has more general integrals:
\begin{eqnarray}\label{lemma1_3}
 \frac{1}{(2\pi \hbar)^{n/2}}\int_{R^{n}}\chi (x-y) e^{\frac{iy(p-k)}{\hbar}}dy
 =\tilde\chi (p-k)e^{\frac{ix(p-k)}{\hbar}},&&\\
 \label{lemma1_4}
\frac{1}{(2\pi \hbar)^{n/2}}\int_{R^{n}}\tilde\chi (p-k)e^{\frac{ik(x-y)}{\hbar}}dk
=\chi (x-y)e^{\frac{ip(x-y)}{\hbar}}.&&
\end{eqnarray}
Functions $\chi $ and $\tilde\chi $ satisfy the following relations:
\begin{equation}\label{lemma1_5}
\chi (\alpha )\chi (\beta )\!=\!\chi \!\left(\!\frac{\alpha \!+\!\beta }{\sqrt{2}}\!\right)
\chi \!\left(\!\frac{\alpha \!-\!\beta }{\sqrt{2}}\!\right)
\ \mbox{and }\
\tilde\chi (\alpha )\tilde\chi (\beta )\!=\!\tilde\chi\!\left(\!\frac{\alpha \!+\!\beta }{\sqrt{2}}\!\right)
\tilde\chi\!\left(\!\frac{\alpha\!-\!\beta }{\sqrt{2}}\!\right)\!.
\end{equation}
The derivatives of the functions $\chi(y)$ and $\tilde\chi(k)$ have the form
\begin{equation}\label{lemma1_diff}
\frac{\partial \chi(y)}{\partial y_j}=-b/(a\hbar)\ y_j \chi(y);\ \ \ \ \ \frac{\partial \tilde\chi'(k)}{\partial k_j}=-a/(b\hbar)\ k_j \tilde\chi(k).
\end{equation}
The functions $\chi^2 $ and $\tilde\chi^2 $  are the densities of
the normal distribution in the configuration space and in the space of momenta,
respectively, with the zero mathematical expectations and the dispersions equal to
$a\hbar/(2b)$ and $b\hbar/(2a)$. I.~e., the following equalities for the function $\chi $ hold:
\begin{eqnarray}\label{lemma1_6}
 \int_{R^{n}}\chi^2 (\eta )d\eta =1;\ \ \ \ \ \ \ \ \ \ \ \ \ \ \
 \int_{R^{n}}\eta_i \chi^2 (\eta )d\eta =0,\ i=1,\ldots,n;\ \ \ &&\nonumber\\
\int_{R^{n}}\eta_i \eta_j  \chi^2 (\eta )d\eta =0, \mbox{ for }i\not =j;\
\int_{R^{n}}\eta_i^2 \chi^2 (\eta )d\eta= a\hbar/(2b),  \ i=1,\ldots,n.&&
\end{eqnarray}
The other moments have order $o(\hbar)$. Analogous equalities hold for
the function $\tilde\chi $:
\begin{eqnarray}\label{lemma1_7}
 \int_{R^{n}}\tilde\chi^2 (\xi )d\xi  =1;\ \ \ \ \ \ \ \ \ \ \ \ \ \ \
 \int_{R^{n}}\xi_i \tilde\chi^2 (\xi )d\xi  =0,\ i=1,\ldots,n;\ \ \ &&\nonumber\\
\int_{R^{n}}\xi_i \xi_j  \tilde\chi^2 (\xi )d\xi  =0, \mbox{ for }i\not =j;\
\int_{R^{n}}\xi_i^2 \tilde\chi^2 (\xi )d\xi = b\hbar/(2a), \ i=1,\ldots,n.&&
\end{eqnarray}
Besides that, the following equalities hold:
\begin{eqnarray}\label{lemma1_8}
 \frac{1}{(2\pi \hbar)^{n/2}}\int_{R^{n}}\eta_i\xi_j
 \chi (\eta )\tilde\chi (\xi )
 e^{\frac{i\eta \xi}{\hbar}}d\eta d\xi =0, \mbox{ for }i\not =j;\ \ \ &&\\
\label{lemma1_9}
 \frac{1}{(2\pi \hbar)^{n/2}}\int_{R^{n}}\eta_j\xi_j
 \chi (\eta )\tilde\chi (\xi )
 e^{\frac{i\eta \xi}{\hbar}}d\eta d\xi =\frac{i\hbar}{2},\ j=1,\ldots,n.&&
\end{eqnarray}
}

All integrals from Lemma~1, except for the latter one (\ref{lemma1_9}), are well known.
The latter integral has been computed by the author
using the system Mathematica 5.0  \cite{math}.

{\bf Lemma 2.} {\it  Let
\begin{equation}\label{lemma2_D}
D= \frac{1}{(2\pi \hbar)^{n}2^{n/2}}\ \chi (\eta )\ \chi\!\left({\eta'}\!/\!{\sqrt{2}}\right)\tilde\chi\!\left({\xi'}\!/\!{\sqrt{2}}\right)\tilde\chi (\xi )
e^{{i(\eta \xi +\eta \xi'+\eta'\xi +\eta'\xi'/2)}/{\hbar}}.
\end{equation}
Then the following equalities hold:
\begin{eqnarray}
 &&1)\  \int_{R^{4n}}Dd\eta\ d\xi\ d\eta' d\xi'=1;\ \ \ \
2)\  \int_{R^{4n}}\eta_j Dd\eta\ d\xi\ d\eta' d\xi'=0,\ \ j=1,\ldots,n;\nonumber\\
&&3)\ \int_{R^{4n}}(\eta_j +\eta'_j) Dd\eta\ d\xi\ d\eta' d\xi'=0,\ \ j=1,\ldots,n; \nonumber\\
&&4)\  \int_{R^{4n}}\xi _j Dd\eta\ d\xi\ d\eta' d\xi'=0,\ \ j=1,\ldots,n;\nonumber\\
&&5)\ \int_{R^{4n}}(\xi _j +\xi'_j) Dd\eta\ d\xi\ d\eta' d\xi'=0,\ \ j=1,\ldots,n; \nonumber\\
&&6)\  \int_{R^{4n}}\eta_i (\eta_j +\eta'_j) Dd\eta\ d\xi\ d\eta' d\xi'=0,\ \ i,j=1,\ldots,n;\nonumber\\
&&7)\ \int_{R^{4n}}(\xi_i +\xi'_i)\xi_j Dd\eta\ d\xi\ d\eta' d\xi'=0,\ \ i,j=1,\ldots,n; \nonumber\\
&&8)\  \int_{R^{4n}}\eta_i \eta_j  Dd\eta\ d\xi\ d\eta' d\xi'=0,\ \ \mbox{for}\ \ i\not=j;\nonumber\\
&&\ \ \ \  \int_{R^{4n}}\eta^2_j   Dd\eta\ d\xi\ d\eta' d\xi'=a\hbar/(2b),\ \ j=1,2\ldots,n;\nonumber\\
&&9)\  \int_{R^{4n}}(\eta_i +\eta'_i)(\eta_j +\eta'_j) Dd\eta\ d\xi\ d\eta' d\xi'=0,
\ \ \mbox{for}\ \ i\not=j;\nonumber\\
&&\ \ \ \ \  \int_{R^{4n}}(\eta_j +\eta'_j)^2 Dd\eta\ d\xi\ d\eta' d\xi'=a\hbar/(2b),
\ \ j=1,2\ldots,n;\nonumber\\
&&10)\  \int_{R^{4n}}\xi_i \xi_j  Dd\eta\ d\xi\ d\eta' d\xi'=0,\ \ \mbox{for}\ \ i\not=j;
\nonumber\\
&&\ \ \ \ \  \int_{R^{4n}}\xi^2_j   Dd\eta\ d\xi\ d\eta' d\xi'=b\hbar/(2a),\ \ j=1,2\ldots,n;
\nonumber\\
&&11)\  \int_{R^{4n}}(\xi_i +\xi'_i)(\xi_j +\xi'_j) Dd\eta\ d\xi\ d\eta' d\xi'=0,\ \ \mbox{for}
\ \ i\not=j;\nonumber\\
&&\ \ \ \ \  \int_{R^{4n}}(\xi_j +\xi'_j)^2 Dd\eta\ d\xi\ d\eta' d\xi'=b\hbar/(2a),\ \
j=1,2\ldots,n;\nonumber\\
&&12)\ \int_{R^{4n}}\eta_i \xi_j Dd\eta\ d\xi\ d\eta' d\xi'=0,\ \ i,j=1,\ldots,n; \nonumber\\
&&13)\ \int_{R^{4n}}\eta'_i \xi'_j Dd\eta\ d\xi\ d\eta' d\xi'=0,\ \ i,j=1,\ldots,n; \nonumber\\
&&14)\  \int_{R^{4n}}\eta_i (\xi_j +\xi'_j)  Dd\eta\ d\xi\ d\eta' d\xi'=0,\ \ \mbox{for}\ \
i\not=j;\nonumber\\
&&\ \ \ \ \  \int_{R^{4n}}\eta_j (\xi_j +\xi'_j) Dd\eta\ d\xi\ d\eta' d\xi'=i\hbar/2,\ \
j=1,2\ldots,n;\nonumber\\
&&15)\  \int_{R^{4n}}\xi _i (\eta_j +\eta'_j)  Dd\eta\ d\xi\ d\eta' d\xi'=0,\ \ \mbox{for}
\ \ i\not=j;\nonumber\\
&&\ \ \ \ \  \int_{R^{4n}}\xi _j (\eta _j +\eta'_j) Dd\eta\ d\xi\ d\eta' d\xi'=i\hbar/2,
\ \ j=1,2\ldots,n;\nonumber\\
&&16)\  \int_{R^{4n}}(\xi _i+\xi' _i) (\eta_j +\eta'_j)  Dd\eta\ d\xi\ d\eta' d\xi'=0,\ \
\mbox{for}\ \ i\not=j;\nonumber\\
&&\ \ \ \ \  \int_{R^{4n}}(\xi _j+\xi' _j) (\eta _j +\eta'_j) Dd\eta\ d\xi\ d\eta' d\xi'=i\hbar,
\ \ j=1,2\ldots,n.\nonumber
\end{eqnarray}
}

{\bf Proof of Lemma 2} is based on the use of formulas from Lemma~1.
Let us compute here the first integral of Lemma~2. The other integrals are computed in a
similar way.

For the first integral, after substitution into it, instead of $D$, of
expression~(\ref{lemma2_D}), we obtain
\begin{eqnarray}
\!\!\!\!&&\!\!\!\!\frac{1}{(2\pi \hbar)^{n}2^{n/2}}\!\! \int_{R^{4n}}\!\!
\chi (\eta )\! \chi\!\left({\eta'}/{\sqrt{2}}\right)\!\tilde\chi\!\left({\xi'}/{\sqrt{2}}\right)
\!\tilde\chi (\xi )
 e^{{i(\eta \xi +\eta \xi'+\eta'\xi +\eta'\xi'/2)}/{\hbar}}d\eta\ d\xi\ d\eta' d\xi'
\nonumber\\
&&
=\{ integrating\ over\ \xi\  by\ formula~(\ref{lemma1_2})  \}
\nonumber\\
&&
=\frac{1}{(2\pi \hbar)^{n/2}2^{n/2}}\!\! \int_{R^{3n}}\!\!
\chi (\eta )
\chi\!\left({\eta'}\!/\!{\sqrt{2}}\right)\!\tilde\chi\!\left({\xi'}\!/\!{\sqrt{2}}\right)
\chi (\eta\! +\!\eta')
e^{{i(\eta  +\eta'/2)\xi'}/{\hbar}}d\eta  d\eta' d\xi'
\nonumber\\
&&
=\{ integrating\ over\ \xi'/\sqrt{2}\  by\ formula~(\ref{lemma1_2})  \}
\nonumber\\
&&
=\int_{R^{2n}}\chi (\eta ) \chi\left({\eta'}/{\sqrt{2}}\right)
\chi\left(\sqrt{2}\eta +{\eta'}/{\sqrt{2}}\right)\chi (\eta +\eta')d\eta  d\eta'
\nonumber\\
&&=\{apply\ formula~(\ref{lemma1_5})\ to\ the\ two\ middle\ factors\ \chi \}
\nonumber\\
&&
=\int_{R^{2n}}\chi^2 (\eta ) \chi ^2(\eta +\eta')d\eta  d\eta'=1 \ \  \{ by\
formula~(\ref{lemma1_6})   \}.
\nonumber
\end{eqnarray}

In the proof of Statement 2 we shall also need the averaging of functions $F(x,p)$
on the phase space $(x,p) \in R^{2n}$
with respect to the density of the distribution
\begin{equation}\label{W'_psi}
 W'_{\psi}=\frac{1}{(2\pi \hbar)^{n}}\int_{R^{n}} \psi(x) \psi^*(y) e^{\frac{i (y-x)p}{\hbar}}dy.
\end{equation}
This density looks similar to Wigner's quasidistribution, but does not coincide with it.

Denote by $\bar F_{W'_{\psi}}$ the average of the function $F(x,p)$ with respect to the
density $W'_{\psi}$. That is,
\begin{equation}\label{F_W'}
\bar F_{W'_{\psi}}=\frac{1}{(2\pi \hbar)^{n}}\int_{R^{3n}} F(x,p) \psi(x) \psi^*(y)
e^{\frac{i (y-x)p}{\hbar}}dy dx dp.
\end{equation}

{\bf Lemma 3.} {\it  If $F(x,p)$ is a smooth function
which, together with all its derivatives, grows at infinity no faster than a polynomial,
and
$\psi(x)$ is an arbitrary smooth complex valued function rapidly decreasing at
infinity, then the following equality holds:
$$
\lim_{\hbar\rightarrow 0}\bar F_{W'_{\psi}}=\int_{R^{n}} F(x,0) \psi(x) \psi^*(x) dx.
$$
}

{\bf Proof of Lemma 3.} Let us make the change of variables $k=p/{\hbar}$
under the integral (\ref{F_W'}), and let us integrate over $y$. We obtain
\begin{eqnarray}\label{F_W'_tilde}
\bar F_{W'_{\psi}}&=&\frac{1}{(2\pi)^{n}}\int_{R^{3n}} F(x,\hbar k) \psi(x) \psi^*(y)
e^{i (y-x)k}dy dx dk\nonumber\\
&=&\frac{1}{(2\pi)^{n/2}}\int_{R^{2n}} F(x,\hbar k) \psi(x) \tilde\psi^*(k) e^{-i xk}dx dk,\\
\mbox{where}\ \ \ \tilde\psi^*(k) &=&\frac{1}{(2\pi)^{n/2}}\int_{R^{n}} \psi^*(y) e^{i yk}dy
\end{eqnarray}
is the Fourier transform of the function $\psi^*(y)$.
Since the function $\psi^*(y) $ is rapidly decreasing, its Fourier transform
$\tilde\psi^*(k)$ is also a function rapidly decreasing at infinity.

Since $\hbar$ is a small quantity, let us decompose the smooth function $F(x,\hbar k)$
over $\hbar k$
by the Taylor formula up to the terms of first order. We have
$$F(x,\hbar k)= F(x,0) + \hbar\sum_{i=1}^{n} k_i \frac{\partial F}{\partial p_i}
(x, \theta \hbar k),$$
$$\mbox{where}\ \ \theta =(\theta_1,...,\theta_n )\ \mbox{and}\ 0\leq \theta_i \leq 1\ \mbox{for}
\  i=1, ..., n.$$
Let us substitute this expression of the function $F(x,\hbar k)$ into (\ref{F_W'_tilde}).
We obtain
\begin{eqnarray}\label{F_W'_tilde_Talor}
\bar F_{W'_{\psi}}\!\!&=&\!\!\frac{1}{(2\pi)^{n/2}}\int_{R^{2n}}
\left ( F(x,0) + \hbar\sum_{i=1}^{n} k_i \frac{\partial F}{\partial p_i}
(x, \theta \hbar k)\right )
\psi (x) \tilde\psi^*(k) e^{-i xk}dx dk\nonumber\\
&=&\frac{1}{(2\pi)^{n/2}}\int_{R^{2n}}F(x,0) \psi(x) \tilde\psi^*(k) e^{-i xk}dx dk\nonumber\\
& &+\frac{\hbar}{(2\pi)^{n/2}}\int_{R^{2n}}\sum_{i=1}^{n} k_i \frac{\partial F}{\partial p_i}
(x, \theta \hbar k)
\psi(x) \tilde\psi^*(k) e^{-i xk}dx dk.
\end{eqnarray}

Let us estimate the coefficient before $\hbar$ in the second summand of the obtained equality.
We have
\begin{eqnarray}
&&\left | \frac{1}{(2\pi)^{n/2}}\int\limits_{R^{2n}}\sum_{i=1}^{n} k_i \frac{\partial F}
{\partial p_i}(x, \theta \hbar k)
\psi(x) \tilde\psi^*(k) e^{-i xk}dx dk \right |\nonumber\\
&&\leq
\frac{1}{(2\pi)^{n/2}}\int\limits_{R^{2n}}\left | \sum_{i=1}^{n} k_i \frac{\partial F}
{\partial p_i}(x, \theta \hbar k) \right |
|\psi(x)|\  |\tilde\psi^*(k)| dx dk\nonumber\\
&&=
\frac{1}{(2\pi)^{n/2}}\ \lim_{r\rightarrow \infty }\int \limits_{D_r=\{ ||(x,k)||\leq r \}}
\left | \sum_{i=1}^{n} k_i \frac{\partial F}{\partial p_i}(x, \theta \hbar k) \right |
|\psi(x)|\  |\tilde\psi^*(k)| dx dk\nonumber\\
&&\leq
\frac{1}{(2\pi)^{n/2}}\ \lim_{r \rightarrow \infty }\int \limits_{D_r} \max
\limits_{(x,k)\in D_r}
\left(\left | \sum_{i=1}^{n} k_i \frac{\partial F}{\partial p_i}(x, \theta \hbar k)
\right |\right )
|\psi(x)|\  |\tilde\psi^*(k)| dx dk
\nonumber\\
&&\leq
 \frac{1}{(2\pi)^{n/2}}\ \lim_{r\rightarrow \infty }\int \limits_{D_r}
 \max\limits_{(x,k)\in D_r}
\left(\left | \sum_{i=1}^{n} k_i \frac{\partial F}{\partial p_i}(x,  \hbar k) \right |\right )
|\psi(x)|\  |\tilde\psi^*(k)| dx dk \nonumber\\
&&=
 \lim_{r\rightarrow \infty }\int \limits_{D_r}O(x^N, k^N) |\psi(x)|\  |\tilde\psi^*(k)| dx dk=M.
\nonumber
\end{eqnarray}
The latter equalities follow from the fact that by the statement of Lemma~3, the expression
in the integral, standing under the operation $\max$,
grows no faster than a polynomial of certain degree $N$ with respect to each variable,
and $|\psi(x)|$ and $|\tilde\psi^*(k)|$ are rapidly decreasing functions
(decreasing at infinity faster than any power), and the limit as $r\rightarrow \infty$
of the integral of a positive rapidly decreasing function exists and is equal to some $M$.

Formula (\ref{F_W'_tilde_Talor})
and the boundedness of the coefficient before $\hbar\rightarrow 0$ in the second summand
of this formula imply that
\begin{equation}
\bar F_{W'_{\psi}}=\frac{1}{(2\pi)^{n/2}}\int_{R^{2n}}F(x,0) \psi(x)
\tilde\psi^*(k) e^{-i xk}dx dk+ O(\hbar).\nonumber
\end{equation}
Hence
\begin{eqnarray}
\lim_{\hbar\rightarrow 0 }\bar F_{W'_{\psi}}&=&\frac{1}{(2\pi)^{n/2}}\int_{R^{2n}}F(x,0)
\psi(x) \tilde\psi^*(k) e^{-i xk}dx dk\nonumber\\
&=& \int_{R^{n}}F(x,0) \psi(x) \psi^*(x) dx. \nonumber
\end{eqnarray}
The latter equality is obtained by computing the integral over $k$. This integral over $k$
is the inverse Fourier transform of the function $\tilde\psi^*(k)$, and it yields the
function $\psi^*(x).$
Lemma~3 is proved.

The distribution $W'\psi$, as Wigner's distribution, is not nonnegative,
and the distribution $\rho_{\psi},$
given by expression~(\ref{rho_psi_chi})  for the wave function $\psi$, is nonnegative.

Denote by $\bar F_{\rho_{\psi}}$ the average of the function $F(x,p)$ with respect to
the distribution $\rho$. That is,
\begin{eqnarray}\label{F_rho}
\!\!\!\!\!\!\!\!\bar F_{\rho_{\psi}}\!\!\!\!\!&=&\!\!\!\!\!\int_{R^{2n}} F(x,p)
\rho_{\psi}(x,p) dx dp\nonumber\\
&=&\!\!\!\!
\frac{1}{(2\pi{\hbar})^{n}}\!\!\!
\int\limits_{R^{4n}}\!\!\!\!F(x,p)\psi(y)\psi^{*}\!(y')\chi(x-y)\chi(x-y')
e^{{{i (y'-y)p}/{\hbar}}}
dy dy' dx dp.
\end{eqnarray}

{\bf Lemma 4.} {\it  If $F(x,p)$ is a smooth function which, together with its derivatives,
grows at infinity no faster than a polynomial, and
$\psi(x)$ is an arbitrary smooth complex valued function rapidly decreasing at
infinity, then the following equality holds:
\begin{equation}\label{bar_F_rho_and_W'}
\bar F_{W'_{\psi}}=\bar F_{\rho_{\psi}}+O(h)=\bar F_{\rho_{\psi}}+o_{\hbar}(1),
\end{equation}
where $o_{\hbar}(1)$ is an infinitely small quantity with respect to $\hbar$.
}

{\bf Proof of Lemma 4.} Consider $\bar F_{\rho_{\psi}}$ given by expression~(\ref{F_rho}).
Let us represent the function $\bar\psi^*(y') $, using composition of the direct and inverse
Fourier transform~(\ref{fur}), in the following form:
$$
\bar\psi^*(y')={\cal F}^{-1}[{\cal F}[\bar\psi^*(y''), y''\rightarrow k], k\rightarrow y'],
$$
i. e., in the following form:
$$
\bar\psi^*(y')=\frac{1}{(2\pi \hbar)^{n}}\int_{R^{2n}}\bar\psi^*(y'')
e^{\frac{i(y''-y')k}{\hbar}}dk\ dy''.
$$
Let us substitute this expression into expression~(\ref{F_rho}).
After simple transformations under integral we obtain
\begin{eqnarray}
\bar F_{\rho_{\psi}}=\frac{1}{(2\pi{\hbar})^{2n}}
\int_{R^{6n}}F(x,p)\psi(y)\psi^{*}\!(y'')\chi(x-y)\chi(x-y') &&\nonumber\\
\times e^{{i (y''k-yp+y'p-y'k)}/{\hbar}}dy' dy dy'' dk dx dp.&&
\end{eqnarray}
In this expression, let us integrate over $y'$ using formula~(\ref{lemma1_3})
from Lemma 1. We obtain
\begin{eqnarray}
\bar F_{\rho_{\psi}}=\frac{1}{(2\pi{\hbar})^{3n/2}}
\int_{R^{5n}}F(x,p)\psi(y)\psi^{*}\!(y'')\chi(x-y)\tilde\chi(p-k) &&\nonumber\\
\times e^{{i (y''k-yp+xp-xk)}/{\hbar}} dy dy'' dk dx dp.&&
\end{eqnarray}
In this expression, let us make change of variables, introducing new variables
$\xi =p-k$ and $\eta =x-y$.
Then, $p=k+\xi $, $x=y+\eta $, and
\begin{eqnarray}
\bar F_{\rho_{\psi}}=\frac{1}{(2\pi{\hbar})^{3n/2}}
\int_{R^{5n}}F(y+\eta, k+\xi )\psi(y)\psi^{*}\!(y'')\chi(\eta )\tilde\chi(\xi )
&&\nonumber\\
\times e^{{i (y''k-yk+\eta \xi)}/{\hbar}} d\eta d\xi  dy dy'' dk.&&
\end{eqnarray}
 Assuming $\eta $ and $\xi $ to be small quantities, let us decompose the function
 $F(y+\eta,k+\xi)$ by the Taylor formula at the point $(y,k)$
up to the terms of second order.
We obtain the following expression, in which the values of the function $F$ and its
derivatives are taken at the point $(y,k)$:
\begin{eqnarray}
F(y+\eta,k+\xi)=F(y,k)\!+\!\sum\limits_{i=1}^{n}(F'_{x_i}\eta_i\!\!\!\!&+&\!\!\!\!
F'_{p_i}\xi_i)\nonumber\\
+\frac{1}{2}\sum\limits_{i,j=1}^{n}\!( F''_{x_i,x_j}\eta_i\eta_j+2F''_{x_i,p_j}\eta_i \xi _j
\!\!\!\!&+&\!\!\!\! F''_{p_i,p_j}\xi_i \xi _j)+o(\eta^2,\xi^2,\eta \xi ).
\end{eqnarray}

Let us substitute this decomposition instead of the function $F(y+\eta,k+\xi)$ into the
latter integral, and let us integrate over the variables $\eta $ and $\xi $
using the formulas written out in Lemma~1. We obtain
\begin{eqnarray}\label{bar_F_with_S}
\bar F_{\rho_{\psi}}\!\!=\!\!\frac{1}{(2\pi{\hbar})^{2n}}\!\!\!
\int\limits_{R^{3n}}\!\!\!
\left[F(y,k)\!+\!\frac{1}{2}\sum\limits_{j=1}^{n}\!\left(\!F''_{x_j,x_j}\frac{a\hbar}{2b}+
F''_{x_j,p_j}(i\hbar)
\!+\! F''_{p_j,p_j}\frac{b\hbar}{2a}\right)\!+\!o(\hbar)\right]
&&\nonumber\\
\times \psi(y)\psi^{*}\!(y'') e^{{i (y''-y)k}/{\hbar}}   dy dy'' dk&&\nonumber\\
=\bar F_{W'_{\psi}}+\hbar\bar S_{W'_{\psi}}+o(\hbar),\ \ \ &&
\end{eqnarray}
where    $S(x,p)=\frac{1}{2}\sum\limits_{j=1}^{n}\!\left(\!F''_{x_j,x_j}({a}/{2b})
+iF''_{x_j,p_j}
\!+\! F''_{p_j,p_j}({b}/{2a})\right)$
is the second summand under the integral in expression~(\ref{bar_F_with_S}) divided
by $\hbar$, and the average values $\bar F_{W'_{\psi}}$
and $\bar S_{W'_{\psi}}$ of the functions $F$ and $S$ with respect to the distribution
$W_{\psi}$ are given by expression~(\ref{F_W'}).

Since by statement of Lemma 4, the function $F(x,p)$ grows at infinity, together with its
derivatives, no faster than a polynomial,
the same property holds for the function $S(x,p)$. Let us apply Lemma~3 to the function
$S(x,p)$. We obtain that $\bar S_{W_{\psi}}$
is bounded as $\hbar\rightarrow 0.$ This and equality~(\ref{bar_F_with_S}) imply the equality
$\bar F_{\rho_{\psi}}=\bar F_{W'_{\psi}}+O(\hbar), $
which is equivalent to the equality required in Lemma~4.

Thus, all the Lemmas necessary for the proof of Statement 2, are proved.
Let us now proceed to the proof of Statement 2 itself.

By Statement~1,
$\alpha_{max}\stackrel{def}{=}
2\cdot  \max_{\bar \varphi_0}|| A \bar \varphi_0- P_0 A\bar\varphi_0||$,
where $A$  is given in our case by expression~(\ref{eq_det}), the projection operator
$P_0$ is given by expression~(\ref{p_0}),
and normalized functions $\bar \varphi_0$ are given by expression~(\ref{view_varphi}),
and the function $\bar \varphi_0 \bar \varphi_0^*$ is the probability distribution
in the physical region of the phase space for the given process.

If the operator $A$ is represented as a sum $A=\sum_{j=1}^{2n+1} A_{j}$, then, by the
property of the norm stating that the norm of the sum of vectors is no greater than the
sum of norms of these vectors, we have
\begin{equation}\label{sum_normA_j}
||A\bar\varphi_0- P_0A\bar\varphi_0|| =  ||\sum\limits_{s=1}^{2n+1}( A_s\bar\varphi_0-P_0 A_s\bar\varphi_0)|| \leq
 \sum\limits_{s=1}^{2n+1}||A_s\bar\varphi_0-P_0 A_s\bar\varphi_0||.
\end{equation}
Hence, for the estimate of the quantity $|| A\bar\varphi_0-P_0 A\bar\varphi_0||$
it suffices to estimate the quantities
$||A_s \bar\varphi_0-P_0 A_s\bar\varphi_0||$, for $s=1,\ldots, 2n+1$.

In our case the operator $A$ is given by expression~(\ref{eq_det}):
\begin{equation}
 A \varphi=\sum_{j=1}^{n}
\biggl(
\frac{\partial H}{\partial x_j} \frac{\partial\varphi}{\partial p_j}-
\frac{\partial H}{\partial p_j} \frac{\partial\varphi}{\partial x_j}
\biggr)
-\frac{i}{\hbar}
\biggl(H-\sum_{j=1}^{n}\frac{\partial H}{\partial p_j}p_j\biggr)\varphi;
\nonumber
\end{equation}
\begin{eqnarray}\label{def_A_s}
\mbox{and } A_{2j-1}=\frac{\partial H}{\partial x_j} \frac{\partial}{\partial p_j},  \ \ \
A_{2j}=- \frac{\partial H}{\partial p_j} \frac{\partial}{\partial x_j}, \ \mbox{ for }
j=1,\ldots,n; \nonumber\\
\mbox{ and } \ \ A_{2n+1}=-\frac{i}{\hbar}f(x,p)=-\frac{i}{\hbar}
\left(H-\sum_{j=1}^{n}\frac{\partial H}{\partial p_j}p_j\right).
\end{eqnarray}

Note also that since $P_0$ is a self-adjoint projection operator, the vectors
$P_0 A_s \bar\varphi_0$ and
$A_s \bar\varphi_0-P_0 A_s \bar\varphi_0$ are orthogonal. The sum of these vectors
equals $A_s \bar\varphi_0$. Hence the following equality holds:
\begin{eqnarray}\label{A_minus_P_0_A}
|| A_s \bar \varphi_0- P_0 A_s \bar \varphi_0||^2=||A_s \bar \varphi_0||^2-||P_0 A_s \bar \varphi_0||^2.
\end{eqnarray}

Let us start estimating these quantities, starting with $s=2n+1$.

{\bf 1. The case $A_{2n+1}=-i/\hbar f(x,p)$.}
We estimate $ ||A_{2n+1} \bar \varphi_0-P_0 A_{2n+1} \bar \varphi_0||^2$,
when the operator $A_{2n+1}$ is the operator of multiplication by the
function $-i/\hbar f(x,p)$ and
\begin{equation}\label{f_x_p}
f(x,p)=H-\sum_{j=1}^{n}\frac{\partial H}{\partial p_j}p_j.
\end{equation}

{\bf 1.1. An estimate of $\ ||A_{2n+1} \bar \varphi_0||^2.\ $}
In this case, for the quantity $||A_{2n+1} \bar \varphi_0||^2
=\langle  A_{2n+1} \bar \varphi_0; A_{2n+1} \bar \varphi_0\rangle$,
after substitution of $i/\hbar f(x,p)$ instead of
$A_{2n+1},$  substitution of expressions~(\ref{view_varphi}) for $\bar\varphi_0$,
and multiplication of both parts of the equality by $\hbar^2$, we have
\begin{eqnarray}\label{normA}
\!\!\!\!\!\!\!\hbar^2||A_{2n+1} \bar \varphi_0||^2 \!\!\!\!\!&=&\!\!\!\!\!
\frac{1}{(2\pi \hbar)^{n}}
\!\!\int_{R^{2n}}\!\!dx dp\ f^2(x,p)\!\!\int_{R^{n}}\!\!dy\ \bar\psi(y)\chi (x-y)
e^{\frac{i(x-y)p}{\hbar}}
\nonumber\\
\!\!\!\!\!&&\!\!\!\!\!\times \int_{R^{n}}\bar\psi^*(y') \chi (x-y')
e^{\frac{i(y'-x)p}{\hbar}} dy'
\nonumber\\
\!\!\!\!\!&=&\!\!\!\!\! \frac{1}{(2\pi \hbar)^{n/2}}\!\!\int_{R^{2n}}\!\!dx dp f^2(x,p) \!\!
\int_{R^{n}}\!\!dy\ \bar\psi(y)\chi (x-y)e^{\frac{i(x-y)p}{\hbar}}I_{int},
\end{eqnarray}
\begin{eqnarray}\label{I_int_def}
\mbox{where }\ \ I_{int}=\frac{1}{(2\pi \hbar)^{n/2}} \int_{R^{n}}\bar\psi^*(y')
\chi (x-y')e^{\frac{i(y'-x)p}{\hbar}} dy'.
\end{eqnarray}

To transform the integral $I_{int}$ in this expression,
let us represent the function $\bar\psi^*(y')$, using composition of the direct and inverse
Fourier transform~(\ref{fur}), in the form
$$
\bar\psi^*(y')
={\mathcal F}^{-1}[{\mathcal F}[\bar\psi^*(y''), y''\rightarrow k], k\rightarrow y'],
$$
i.~e., in the form
$$
\bar\psi^*(y')
=\frac{1}{(2\pi \hbar)^{n}}\int_{R^{2n}}\bar\psi^*(y'') e^{\frac{i(y''-y')k}{\hbar}}dk\ dy''.
$$
Let us substitute this expression into expression~(\ref{I_int_def}) for $I_{int}$. We obtain
\begin{eqnarray}\label{I_int}
I_{int}&=&\frac{1}{(2\pi \hbar)^{3n/2}}\int_{R^{3n}} \bar\psi^*(y'')
e^{\frac{i(y''-y')k}{\hbar}}\chi (x-y')e^{\frac{i(y'-x)p}{\hbar}} dy' dk dy''
\nonumber\\
&=&\frac{1}{(2\pi \hbar)^{3n/2}}\int\limits_{R^{2n}} \bar\psi^*(y'')
e^{\frac{i(y''k-xp)}{\hbar}}\int\limits_{R^{n}}\chi (x-y')e^{\frac{iy'(p-k)}{\hbar}}
dy' dk dy''.
\end{eqnarray}
Let us compute the latter integral over $y'$ by formula~(\ref{lemma1_3}) from Lemma~1.
After substitution we obtain
\begin{eqnarray}\label{I_int_res}
I_{int}&=&\frac{1}{(2\pi \hbar)^{n}}\int_{R^{2n}}
\bar\psi^*(y'') e^{\frac{i(y''k-xp)}{\hbar}}\tilde\chi (p-k)e^{\frac{ix(p-k)}{\hbar}}dk dy''
\nonumber\\
&=&\frac{1}{(2\pi \hbar)^{n}}\int_{R^{2n}} \bar\psi^*(y'')
\tilde\chi (p-k) e^{\frac{i(y''k-xk)}{\hbar}}dk dy''.
\end{eqnarray}

Let us substitute the obtained expression for $I_{int}$ into expression~(\ref{normA})
for $\hbar^2||A_{2n+1} \bar \varphi_0||^2$.
After simple transformations we obtain
\begin{eqnarray}
\hbar^2||A_{2n+1} \bar \varphi_0||^2=
\frac{1}{(2\pi \hbar)^{3n/2}}\int_{R^{5n}} f^2(x,p)\bar\psi(y)\bar\psi^*(y'') \chi (x-y)
\tilde \chi (p-k) &&
\nonumber\\
\times
e^{\frac{i(y''k-yp+xp-xk)}{\hbar}} dk\ dy''\ dy\ dx\ dp.&&
\end{eqnarray}
In the obtained expression, let us make change of variables, introducing new variables
$\xi =p-k$ and $\eta =x-y$. Then $p=k+\xi$, $x=y+\eta$, and
\begin{eqnarray}
\hbar^2||A_{2n+1} \bar \varphi_0||^2=
\frac{1}{(2\pi \hbar)^{3n/2}}\int_{R^{5n}}
f^2(y+\eta,k+\xi)\bar\psi(y)\bar\psi^*(y'') \chi (\eta ) \tilde \chi (\xi ) &&
\nonumber\\
\times
e^{\frac{i(y''k-yk+\eta \xi )}{\hbar}} dk\ dy''\ dy\ d\eta \ d\xi.&&
\end{eqnarray}

Assuming that $\eta $ and $\xi $  are small quantities, let us decompose the
function $f^2(y+\eta,k+\xi)$ into the Taylor series at the point $(y,k)$ up to terms of
the second order. We obtain the following expression, in which the values of the function
$f$ and its derivatives are taken at the point $(y,k)$,
\begin{eqnarray}
f^2(y+\eta,k+\xi)\!\!\!\!&=&\!\!\!\!f^2\!+\!\sum\limits_{i=1}^{n}(2f f'_{x_i}\eta_i\!
+\!2f f'_{p_i}\xi_i)\! +\!
\sum\limits_{i,j=1}^{n}\![(f'_{x_j} f'_{x_i}\!+\! f f''_{x_i,x_j})\eta_i\eta_j\!
\nonumber\\
&&+
(f'_{x_j} f'_{p_i}+f f''_{p_i,x_j})\xi_i\eta_j+
(f'_{p_j} f'_{x_i}+ f f''_{x_i,p_j})\eta_i \xi _j
\nonumber\\
&&+
(f'_{p_j} f'_{p_i}+f f''_{p_i,p_j})\xi_i \xi _j]+o(\eta^2,\xi^2,\eta \xi ).
\end{eqnarray}

Let us substitute this expression, instead of function $f^2(y+\eta,k+\xi)$, into the
latter integral and let us perform integration over the variables
$\eta $ and $\xi $ using the formulas written out in Lemma~1. We obtain
\begin{eqnarray}\label{normA_res}
\hbar^2||A_{2n+1} \bar \varphi_0||^2&=&
\frac{1}{(2\pi \hbar)^{n}}\int_{R^{3n}}
\biggl [f^2+
\sum\limits_{j=1}^{n}\biggl[(f'_{x_j} f'_{x_j}+ f f''_{x_j,x_j})\frac{a\hbar}{2b}
\nonumber\\
&&
+
(f'_{x_j} f'_{p_j}+f f''_{p_j,x_j})i\hbar
+(f'_{p_j} f'_{p_j}+f f''_{p_j,p_j})\frac{b\hbar}{2a}\biggr]+o(\hbar)\biggr]
\nonumber\\
&&\times
\bar\psi(y)\bar\psi^*(y'')e^{\frac{i(y''k-yk )}{\hbar}} dk\ dy''\ dy.
\end{eqnarray}

{\bf 1.2. An estimate of $||P_0 A_{2n+1} \bar \varphi_0||^2$.} Let us now estimate
the expression subtracted in~(\ref{A_minus_P_0_A}), i.~e.,
$||P_0 A_{2n+1} \bar \varphi_0||^2=
\langle  P_0 A_{2n+1} \bar \varphi_0; A_{2n+1} \bar \varphi_0\rangle$, where
the operator $A_{2n+1}$ is the multiplication by the smooth function $-i/\hbar f(x,p)$.
Expanding this expression with the scalar product and substituting into it
the expression for the operator $A_{2n+1}$,
expression~(\ref{view_varphi}) for $\bar\varphi_0$ and expression~(\ref{p_0})
for the operator $P_0$, represented, in the notations of Lemma~1
in the form
$$
P_0 \varphi =\frac{1}{(2\pi \hbar)^{n/2}2^{n/2}}
\int_{R^{2n}}\varphi (x',p')\chi \biggl(\frac{x-x'}{\sqrt{2}}\biggr)
\tilde\chi \biggl(\frac{p'-p}{\sqrt{2}}\biggr)e^{\frac{i(x-x')(p+p')}{2\hbar}}dx'dp',
$$
and multiplying both parts of the equality by $\hbar^2$, we obtain
\begin{eqnarray}\label{normP_0A}
\hbar^2 ||P_0 A_{2n+1} \bar \varphi_0||^2&
=&\frac{1}{(2\pi \hbar)^{3n/2}2^{n/2}}\int_{R^{6n}}\bar\psi(y)\chi (x'-y)
e^{\frac{i(x'-y)p'}{\hbar}}f(x',p') \nonumber\\
&&\times \chi \biggl(\frac{x-x'}{\sqrt{2}}\biggr)\tilde\chi
\biggl(\frac{p'-p}{\sqrt{2}}\biggr)e^{\frac{i(x-x')(p+p')}{2\hbar}}f(x,p)\nonumber\\
&& \times  \bar\psi^*(y') \chi (x-y')
e^{\frac{i(y'-x)p}{\hbar}}dy'\ dy\ dx'\ dp'\ dx\ dp\nonumber\\
&=&\frac{1}{(2\pi \hbar)^{n}2^{n/2}}\int_{R^{5n}}\bar\psi(y)\chi (x'-y)
e^{\frac{i(x'-y)p'}{\hbar}}f(x',p') \nonumber\\
&&\times \chi \biggl(\frac{x-x'}{\sqrt{2}}\biggr)\tilde\chi
\biggl(\frac{p'-p}{\sqrt{2}}\biggr)e^{\frac{i(x-x')(p+p')}{2\hbar}} \nonumber\\
&& \times f(x,p)\ I_{int}\ dy\ dx'\ dp'\ dx\ dp,
\end{eqnarray}
\begin{eqnarray}
\mbox{where }\ \ I_{int}=\frac{1}{(2\pi \hbar)^{n/2}}
\int_{R^{n}}\bar\psi^*(y') \chi (x-y')e^{\frac{i(y'-x)p}{\hbar}} dy'.
\nonumber
\end{eqnarray}
The integral $I_{int}$ is the same as in (\ref{I_int_def}) in the computation of
$\hbar^2||A_{2n+1} \bar \varphi_0||^2 $. Let us substitute into expression~(\ref{normP_0A})
the representation of the integral $I_{int}$ in the form~(\ref{I_int_res}).
After simple transformations we obtain,
\begin{eqnarray}
\!\!\!\!\hbar^2 ||P_0 A_{2n+1} \bar \varphi_0||^2\!\!\!\!&=&
\!\!\!\!\frac{1}{(2\pi \hbar)^{2n}2^{n/2}}\int_{R^{7n}} f(x',p')
f(x,p) \bar\psi(y)\bar\psi^*(y'')
\nonumber\\
&&\!\!\!\!\times \chi (x'-y) \chi \biggl(\frac{x-x'}{\sqrt{2}}\biggr)
\tilde\chi \biggl(\frac{p'-p}{\sqrt{2}}\biggr)\tilde\chi (p-k)
\nonumber\\
&&\!\!\!\!\times
e^{\frac{i(x-x')(p+p')}{2\hbar}+\frac{i(y''k-yp'+x'p'-xk)}{\hbar}}dk\ dy'' dy\ dx' dp' dx\ dp.
\end{eqnarray}

In the obtained integral, let us make a change of variables, introducing the new variables
$
\eta =x'-y,\ \ \ \xi =p-k,\ \ \ \eta'=x-x',\ \ \ \xi' =p'-p.
$

Then,
$
x'=y+\eta,\ \ \ p=k+\xi ,\ \ \ x=y+\eta +\eta',\ \ \ p'=k+\xi +\xi'.
$

After substitution of $x', x, p, p'$, expressed through the new variables, and after
simple transformations, we obtain
\begin{eqnarray}\label{normP_0A1}
\hbar^2 ||P_0 A_{2n+1} \bar \varphi_0||^2\!\!\!&=&\!\!\!
\frac{1}{(2\pi \hbar)^{2n}2^{n/2}}\int_{R^{7n}} f(y\!+\!\eta,k\!+\!\xi\!+\!\xi')
f(y\!+\!\eta \!+\!\eta',k\!+\!\xi)
\nonumber\\
&&\!\!\!\times\bar\psi(y)\bar\psi^*(y'')
\chi (\eta ) \chi\left({\eta'}/{\sqrt{2}}\right)\tilde\chi\left({\xi'}/{\sqrt{2}}\right)
\tilde\chi (\xi )
\nonumber\\
&&\!\!\!\times
e^{\frac{i(y''-y)k}{\hbar}+\frac{i(\eta \xi +\eta \xi'+\eta'\xi )}
{\hbar}+\frac{i\eta'\xi'}{2\hbar}}dk\ dy'' dy\ d\eta \ d\xi \ d\eta' d\xi'.
\end{eqnarray}

Since the functions $\chi^2$ and $\tilde\chi^2$ yield the densities of normal distributions
with small dispersions, let us decompose the function
$f(y+\eta,k+\xi +\xi') f(y+\eta +\eta',k+\xi) $ in the previous expression into the Taylor
series at the point $(y,k)$ up to terms of the second order,
assuming the quantities $\eta$, $\eta'$, $\xi$, $\xi'$ to be small. We have,
\begin{eqnarray}\label{ff_ryad}
&&f(y+\eta,k+\xi +\xi') f(y+\eta +\eta',k+\xi)\nonumber\\
&& \ \ \ =f^2(y,k)+\sum\limits_{j=1}^{n}
[f f'_{x_j}\eta_j +ff'_{p_j}(\xi_j +\xi'_j)+ff'_{x_j}(\eta_j +\eta'_j)+ff'_{p_j}\xi_j]
\nonumber\\
\!\!\!\!\!\!\!\!&&\!\!\!\!\!\!\!\!\!+\!\sum\limits_{i,j=1}^{n}\!
[f'_{x_i}f'_{x_j}\eta_i (\eta_j \!+\!\eta'_j)\!+\!f'_{x_i}f'_{p_j}\eta_i \xi_j\! +
\!f'_{p_i}f'_{x_j}(\xi_i\!+\!\xi'_i)(\eta_j \!+\!\eta'_j)\!+\!
                            f'_{p_i}f'_{p_j}(\xi_i\!+\!\xi'_i)\xi_j ]\!
\nonumber\\
\!\!\!\!\!\!\!\!&&\!\!\!\!\!\!\!\!\!+\frac{1}{2}\!\sum\limits_{i,j=1}^{n}\!f
[f''_{x_i,x_j}\eta_i\eta_j\!+\!f''_ {x_i,p_j}\eta_i(\xi_j\! +\!\xi'_j)\!+
\!f''_ {p_i,x_j}(\xi _i\!+\!\xi'_i)\eta_j \!+\!
f''_{p_i,p_j}(\xi _i\!+\!\xi'_i)(\xi _j\!+\!\xi'_j) ]\!
\nonumber\\
\!\!\!\!\!\!\!\!&&\!\!\!\!\!\!\!\!\!+\frac{1}{2}\!\sum\limits_{i,j=1}^{n}\!f[f''_{x_i,x_j}
(\eta_i\!+\!\eta'_i)(\eta_j\!+\!\eta'_j)\!+\!f''_ {x_i,p_j}(\eta_i\!+\!\eta' _i)\xi_j\!+
\!f''_ {p_i,x_j}\xi _i(\eta_j \!+\!\eta'_j)\!+\!
                                                f''_{p_i,p_j}\xi _i \xi _j ]\!
\nonumber\\
&&+o\left(\eta^2, (\eta+\eta')^2,  \xi^2, (\xi +\xi')^2, \eta (\xi +\xi'), (\eta+\eta')\xi
\right).
\end{eqnarray}
After substitution of the decomposition of the function
$f(y+\eta,k+\xi +\xi') f(y+\eta +\eta',k+\xi) $ in the form~(\ref{ff_ryad}) into
expression~(\ref{normP_0A1}) and computing integrals over $\eta$, $\eta'$, $\xi$, $\xi'$ using
the integrals of Lemma~2, we obtain

\begin{eqnarray}\label{normP_0A_res}
\hbar^2 ||P_0 A_{2n+1} \bar \varphi_0||^2\!\!\!&=&\!\!\!\frac{1}{(2\pi \hbar)^{n}}
\int_{R^{3n}}[f^2+\sum\limits_{j=1}^n
[f f''_{x_j,x_j}a \hbar/(2b)\nonumber\\
&&\!\!\!+ (f'_{x_j}f'_{p_j}+f f''_{x_j,p_j}) i \hbar +f f''_{p_j,p_j} b \hbar/(2a)]
+o(\hbar)]\nonumber\\
&&\!\!\!\times \bar\psi(y)\bar \psi^*(y'') e^{\frac{i (y''-y)k}{\hbar}}dk\ dy'' dy.
\end{eqnarray}

Thus, we have estimated the expression $\hbar^2 ||A_{2n+1} \bar \varphi_0||^2$
by formula~(\ref{normA_res}) and expression $\hbar^2 ||P_0 A_{2n+1} \bar \varphi_0||^2$
by formula~(\ref{normP_0A_res}). Let us substitute these formulas into
expression~(\ref{A_minus_P_0_A}), let us reduce similar terms, and let us divide both
parts of the equality by $\hbar^2$.
We obtain
\begin{eqnarray}\label{normP_0A_minus_A_res}
 ||A_{2n+1} \bar \varphi_0-P_0 A_{2n+1} \bar \varphi_0||^2=||A_{2n+1}\bar \varphi_0||^2-
 ||P_0 A_{2n+1} \bar \varphi_0||^2
\nonumber\\
=\frac{1}{\hbar}\frac{1}{(2\pi \hbar)^{n}}\int_{R^{3n}}\biggl [
\sum\limits_{j=1}^{n}\biggl((f'_{x_j})^2\frac{a}{2b}+(f'_{p_j})^2\frac{b}{2a}\biggr)
+o_{\hbar}(1)\biggr]
\nonumber\\
\times  \bar\psi(y)\bar\psi^*(y'')e^{\frac{i(y''k-yk )}{\hbar}} dk\ dy''\ dy.
\end{eqnarray}

The last row of expression~(\ref{normP_0A_minus_A_res}) is the density of the distribution
$W'_{\bar\psi}$ given by expression~(\ref{W'_psi}). Let us apply Lemma~4 to the right hand side
of the equality~(\ref{normP_0A_minus_A_res}). We obtain
\begin{eqnarray}\label{normP_0A_minus_A_res_rho}
 ||A_{2n+1} \bar \varphi_0\!-\!P_0 A_{2n+1} \bar \varphi_0||^2\!=
 \!\frac{1}{\hbar}\biggl [\int\limits_{R^{2n}}\!
\sum\limits_{j=1}^{n}\biggl(\!(f'_{x_j})^2\frac{a}{2b}\!+\!(f'_{p_j})^2\frac{b}{2a}
\!\biggr)\rho_{\bar\psi}dó  dk\!+\!o_{\hbar}(1)\biggr]\!,
\end{eqnarray}
where $\rho_{\bar\psi}(ó,k)$ is the nonnegative density of distribution
given by expression~(\ref{rho_psi_chi})  for the function
$\bar\psi(ó)$.

To estimate the required expression
$\max_{\bar \varphi_0} ||A_{2n+1} \bar \varphi_0-P_0 A_{2n+1} \bar \varphi_0||$,
introduce the constant $C_{2n+1}$ by the following equality:
\begin{eqnarray}\label{C_2n_1}
C_{2n+1}^2\stackrel{def}{=}\max\limits_{(x,p)\in U}
\sum\limits_{j=1}^{n}\biggl((f'_{x_j})^2\frac{a}{2b}+(f'_{p_j})^2\frac{b}{2a}\biggr),
\end{eqnarray}
where the maximum is taken over the physical domain $U$ of values of the
coordinates and momenta for the given process,
which contains the support of the density function of the probability distribution
$\rho_{\bar\psi}$.
Then, equality~(\ref{normP_0A_minus_A_res_rho}) implies that
\begin{eqnarray}\label{normP_0A_minus_A_leq}
 ||A_{2n+1} \bar \varphi_0-P_0 A_{2n+1} \bar \varphi_0||^2\leq
\frac{1}{\hbar}\left(\int\limits_{R^{n}}C_{2n+1}^2\rho_{\bar\psi}dy dk+o_{\hbar}(1)\right)
\nonumber\\
=\left(C_{2n+1}^2+o_{\hbar}(1)\right) \frac{1}{\hbar}.
\end{eqnarray}
The latter equality is obtained from the normalization condition for the density
$\rho_{\bar\psi}$, i.~e., from the equality
 $\int_{R^{n}}  \rho_{\bar\psi} dy dk=1$.

Hence, taking the square root from both parts of the inequality, we finally obtain
\begin{eqnarray}\label{res_2n_1}
 ||A_{2n+1} \bar \varphi_0-P_0 A_{2n+1} \bar \varphi_0||\leq
\left(C_{2n+1}+o_{\hbar}(1)\right) \frac{1}{\sqrt {\hbar}}.
\end{eqnarray}
This finishes examining Case 1.

{\bf 2. Case $A_{2j-1}=\frac{\partial H}{\partial x_j} \frac{\partial}{\partial p_j}$.} Let us
estimate the quantity
$||A_{2j-1} \bar \varphi_0-P_0 A_{2j-1} \bar \varphi_0||^2
=||A_{2j-1} \bar \varphi_0||^2-||P_0 A_{2j-1} \bar \varphi_0||^2$
by the same scheme as in the previous case: we separately estimate
$||A_{2j-1} \bar \varphi_0||^2$  and $||P_0 A_{2j-1} \bar \varphi_0||^2$.

{\bf 2.1. An estimate of $||A_{2j-1} \bar \varphi_0||^2$.} Let us substitute into
$||A_{2j-1} \bar \varphi_0||^2
=\langle A_{2j-1} \bar \varphi_0; A_{2j-1} \bar \varphi_0 \rangle$
expression~(\ref{view_varphi})
for $\bar\varphi_0$, and put
$A_{2j-1}=\frac{\partial H}{\partial x_j} \frac{\partial}{\partial p_j}$. We have
\begin{eqnarray}\label{normA_2j_1}
||A_{2j-1} \bar \varphi_0||^2=\int_{R^{2n}}\frac{\partial H}{\partial x_j} \frac{\partial}
{\partial p_j}
\left(\frac{1}{(2\pi \hbar)^{n/2}}\int_{R^{n}} \bar\psi(y)\chi (x-y) e^{\frac{i(x-y)p}{\hbar}}
dy\right)
\nonumber\\
\times
\frac{\partial H}{\partial x_j} \frac{\partial}{\partial p_j}
\left(\frac{1}{(2\pi \hbar)^{n/2}}\int_{R^{n}} \bar\psi^*(y')\chi (x-y')
 e^{\frac{i(y'-x)p}{\hbar}}dy'\right)
dx dp
\nonumber\\
= \int_{R^{2n}}\frac{\partial H}{\partial x_j}
\left(\frac{1}{(2\pi \hbar)^{n/2}}\int_{R^{n}}
\bar\psi(y)\chi (x-y) \frac{i(x_j-y_j)}{\hbar}e^{\frac{i(x-y)p}{\hbar}}dy\right)
\nonumber\\
\times
\frac{\partial H}{\partial x_j} \frac{\partial}{\partial p_j}\left(I_{int}\right)
dx dp,
\end{eqnarray}
where $I_{int}$ is given by expression~(\ref{I_int_def}).  Let us substitute here,
instead of $I_{int}$, its expression in the form~(\ref{I_int_res}).
After simple transformations and after substitution of the derivative
of the function $\tilde\chi$ in the form~(\ref{lemma1_diff}), we obtain
\begin{eqnarray}
||A_{2j-1} \bar \varphi_0||^2=
 \frac{1}{(2\pi \hbar)^{3n/2}}\int_{R^{5n}}\frac{\partial H}{\partial x_j}
\left( \bar\psi(y)\chi (x-y) \frac{i(x_j-y_j)}{\hbar}e^{\frac{i(x-y)p}{\hbar}}\right)
\nonumber\\
\times
\bar\psi^*(y'')\frac{\partial H}{\partial x_j}
\frac{\partial}{\partial p_j}\left(\tilde\chi(p-k) \right)e^{\frac{i(y''-x)k}{\hbar}}
dk\ dy''dy\ dx\ dp
\nonumber\\
 =- \frac{i a}{b\hbar^2}\frac{1}{(2\pi \hbar)^{3n/2}}
 \int_{R^{5n}}\left(\frac{\partial H}{\partial x_j}\right)^2  \bar\psi(y)\bar\psi^* (y'')
(x_j-y_j)\chi (x-y)
\nonumber\\
\times(p_j-k_j)\tilde\chi(p-k) e^{\frac{i(y''k-yp+xp-xk)}{\hbar}}
dk\ dy''dy\ dx\ dp.
\nonumber
\end{eqnarray}
In the obtained expression, let us make the change of variables
$\eta =x-y$ and $\xi =p-k$. Then, $x=y+\eta $, $p=k+\xi $, and
\begin{eqnarray}
||A_{2j-1} \bar \varphi_0||^2=
- \frac{i a}{b\hbar^2}\frac{1}{(2\pi \hbar)^{3n/2}}\int_{R^{5n}}\left(\frac{\partial H}{\partial x_j}\right)^2  \bar\psi(y)\bar\psi^* (y'')
\eta_j \chi (\eta)
\nonumber\\
\times(\xi_j)\tilde\chi(\xi ) e^{\frac{i(y''-y)k}{\hbar}+\frac{i \eta \xi }{\hbar}}
dk\ dy''dy\ d\eta \ d\xi,
\nonumber
\end{eqnarray}
where in the function $\left(\frac{\partial H}{\partial x_j}\right)^2$,
instead of variables $x$ and $p$, we have substituted $y+\eta$ and $k+\xi$,
respectively.

Assuming $\eta $ and $\xi$ to be small quantities, let us decompose the function
$\left(\frac{\partial H}{\partial x_j}\right)^2(y+\eta, k+\xi)$
into the Taylor series up to the zero order. We have
$$
\left(\frac{\partial H}{\partial x_j}\right)^2\!\!(y+\eta, k+\xi)=
\left(\frac{\partial H}{\partial x_j}\right)^2\!\!(y, k)+O(\eta, \xi).
$$
Let us substitute this expression into the previous one, and integrate it over $\eta$ and $\xi$,
using the integral~(\ref{lemma1_9}) from Lemma~1.
Finally we obtain
\begin{eqnarray}\label{normA_2j_1_res}
||A_{2j-1} \bar \varphi_0||^2=
 \frac{a}{2 b \hbar}\frac{1}{(2\pi \hbar)^{n}}
 \int_{R^{3n}}\biggl(\biggl(\frac{\partial H}{\partial x_j}\biggr)^2+o_{\hbar}(1)\biggr)
 \bar\psi(y)\bar\psi^* (y'')
\nonumber\\
\times e^{\frac{i(y''-y)k}{\hbar}}
dk\ dy''dy.
\end{eqnarray}

{\bf 2.2. An estimate of $||P_0 A_{2j-1} \bar \varphi_0||^2$.} By construction, this
expression satisfies the inequalities
$$
0\leq ||P_0 A_{2j-1} \bar \varphi_0||^2\leq || A_{2j-1} \bar \varphi_0||^2.
$$

Hence and from relations~(\ref{A_minus_P_0_A}) and (\ref{normA_2j_1_res}) we obtain
\begin{eqnarray}\label{normA_2j_1_minus_P0A_res}
||A_{2j-1} \bar \varphi_0\!-\!P_0A_{2j-1} \bar \varphi_0||^2\!=\!
||A_{2j-1} \bar \varphi_0||^2\!-\!||P_0A_{2j-1} \bar \varphi_0||^2\!\leq\! ||A_{2j-1}
 \bar \varphi_0||^2\!
 \nonumber\\
=\frac{a}{2 b \hbar}\frac{1}{(2\pi \hbar)^{n}}
\int_{R^{3n}}\biggl(\biggl(\frac{\partial H}{\partial x_j}\biggr)^2+o_{\hbar}(1)\biggr)
 \bar\psi(y)\bar\psi^* (y'')
\nonumber\\
\times e^{\frac{i(y''-y)k}{\hbar}}
dk\ dy''dy.
\end{eqnarray}
The last row of expression~(\ref{normA_2j_1_minus_P0A_res}) is the density of the
distribution $W'_{\bar\psi}$ given by expression~(\ref{W'_psi}). Let us apply
Lemma~4 to the right hand side of equality~(\ref{normA_2j_1_minus_P0A_res}). We obtain
\begin{eqnarray}\label{normA_2j_1_minus_P0A_res_rho}
 ||A_{2j-1} \bar \varphi_0\!-\!P_0A_{2j-1} \bar \varphi_0||^2\!=\!\frac{a}{2 b \hbar}\biggl
 [\int_{R^{2n}}\!
\biggl(\frac{\partial H}{\partial x_j}\biggr)^2\rho_{\bar\psi}\ dó  dk\!
+\!o_{\hbar}(1)\biggr]\!,
\end{eqnarray}
where $\rho_{\bar\psi}(ó,k)$ is the nonnegative density of distribution given by
expression~(\ref{rho_psi_chi})  for the function $\bar\psi(ó)$.

Introduce the constant $C_{2j-1}$ by the following equality:
\begin{eqnarray}\label{C_2j_1}
C_{2j-1}^2\stackrel{def}{=}\frac{a}{2b}\max\limits_{(x,p)\in U}
\biggl(\frac{\partial H}{\partial x_j}\biggr)^2,
\end{eqnarray}
where the maximum is taken over the physical region $U$ of values of coordinates
and momenta for the given process.

Taking into account this notation, inequality~(\ref{normA_2j_1_minus_P0A_res_rho}) implies that
\begin{eqnarray}
 ||A_{2j-1} \bar \varphi_0-P_0 A_{2j-1} \bar \varphi_0||^2\leq
\frac{1}{\hbar}\left(\int_{R^{2n}}C_{2j-1}^2\rho_{\bar\psi}\ dy dk+o_{\hbar}(1)\right)
\nonumber\\
=\left(C_{2j-1}^2+o_{\hbar}(1)\right) \frac{1}{\hbar}.
\nonumber
\end{eqnarray}
The latter equality is obtained, as in the first case, from the normalization
condition for the density of the probability distribution
$\rho_{\bar\psi}$.

Hence, taking the square root from both parts of the latter inequality, we finally
obtain
\begin{eqnarray}\label{res_2j_1}
 ||A_{2j-1} \bar \varphi_0-P_0 A_{2j-1} \bar \varphi_0||\leq
\left(C_{2j-1}+o_{\hbar}(1)\right) \frac{1}{\sqrt {\hbar}}.
\end{eqnarray}
This finishes examining Case 2.

{\bf 3. Case $A_{2j}=-\frac{\partial H}{\partial p_j} \frac{\partial}{\partial x_j}$.} Let us
estimate the quantity
$||A_{2j} \bar \varphi_0-P_0 A_{2j} \bar \varphi_0||$.

If one applies the operator
$A_{2j}=-\frac{\partial H}{\partial p_j} \frac{\partial}{\partial x_j}$ to the function
$\bar \varphi_0$ of the form~(\ref{view_varphi}), i.~e.,
to the function
$$
\bar\varphi_0(x,p)=\frac{1}{(2\pi{\hbar})^{n/2}}\!\int\limits_{R^n}\!\!\bar\psi(y)\chi(x-y)
e^{{{i  (x-y)p}/{\hbar}}}
dy,
$$
then, taking into account formula~(\ref{lemma1_diff}) for the derivative of the function
$\chi $, we obtain the following equalities:
\begin{eqnarray}
A_{2j}\bar\varphi_0&=&-\frac{\partial H}{\partial p_j }
\frac{\partial \bar\varphi_0}{\partial x_j}=
-\frac{\partial H}{\partial p_j }\frac{1}{(2\pi{\hbar})^{n/2}}\!
\int\limits_{R^n}\!\!\bar\psi(y)
\nonumber\\
&&\times \biggl(-\frac{b}{a\hbar}(x_j-y_j)+
\frac{ip_j}{\hbar}\biggr)\chi(x-y) e^{{{i  (x-y)p}/{\hbar}}}
dy
\nonumber\\
&=&
\frac{b}{a\hbar}\frac{\partial H}{\partial p_j }
\frac{1}{(2\pi{\hbar})^{n/2}}\!\int\limits_{R^n}\!\!\bar\psi(y)
(x_j-y_j)\chi(x-y) e^{{{i  (x-y)p}/{\hbar}}}dy
\nonumber\\
&&-\frac{i p_j}{\hbar}\frac{\partial H}{\partial p_j }
\frac{1}{(2\pi{\hbar})^{n/2}}\!\int\limits_{R^n}\!\!\bar\psi(y)
\chi(x-y) e^{{{i  (x-y)p}/{\hbar}}}dy
\nonumber\\
&=&\frac{-ib}{a}\frac{\partial H}{\partial p_j }
\frac{\partial \bar\varphi_0}{\partial p_j}
+\frac{-i p_j}{\hbar}\frac{\partial H}{\partial p_j }\bar\varphi_0.
\nonumber
\end{eqnarray}
That is,
$A_{2j}\bar\varphi_0=A'_{2j}\bar\varphi_0+A''_{2j}\bar\varphi_0$, where
$$
A'_{2j}\bar\varphi_0\stackrel{def}{=}
-\frac{ib}{a}\frac{\partial H}{\partial p_j } \frac{\partial \bar\varphi_0}{\partial p_j},
\ \ \ \ \ \
A''_{2j}\bar\varphi_0\stackrel{def}{=}
-\frac{i p_j}{\hbar}\frac{\partial H}{\partial p_j }\bar\varphi_0.
$$
Hence, using the property of the norm that the norm of a sum of vectors is no greater
than the sum of norms of these vectors, we obtain the inequality
\begin{eqnarray}\label{A_2j_leq}
||A_{2j} \bar \varphi_0-P_0 A_{2j} \bar \varphi_0||\leq
||A'_{2j} \bar \varphi_0-P_0 A'_{2j} \bar \varphi_0||+||A''_{2j}
\bar \varphi_0-P_0 A''_{2j} \bar \varphi_0||.
\end{eqnarray}

Note that the operator $A'_{2j}$ coincides with the operator $A_{2j-1}$ (see the previous
Case) if one replaces the function $\frac{\partial H}{\partial x_j }$ in it to
$\frac{-ib}{a}\frac{\partial H}{\partial p_j }$. Therefore, using
formulas~(\ref{res_2j_1}) and (\ref{C_2j_1}), we obtain
\begin{eqnarray}\label{res'_2j}
 ||A'_{2j} \bar \varphi_0-P_0 A'_{2j} \bar \varphi_0||\leq
\left(C'_{2j}+o_{\hbar}(1)\right) /{\sqrt {\hbar}},\\
\mbox{where }\ (C'_{2j})^2 \stackrel{def}{=}
\frac{b}{2a}\max\limits_{(x,p)\in U} \biggl(\frac{\partial H}{\partial p_j}\biggr)^2.
\nonumber
\end{eqnarray}

On the other hand, the operator $A''_{2j}$ coincides with the operator $A_{2n+1}$
(see Case~1), in which
$f(x, p)= p_j\frac{\partial H}{\partial p_j}.$
Hence, using formulas~(\ref{res_2n_1}) and (\ref{C_2n_1}), we obtain
\begin{eqnarray}\label{res''_2j}
 ||A''_{2j} \bar \varphi_0-P_0 A''_{2j} \bar \varphi_0||\leq
\left(C''_{2j}+o_{\hbar}(1)\right)/{\sqrt {\hbar}},
\end{eqnarray}
\begin{eqnarray}
\mbox{where }\ (C''_{2j})^2&\stackrel{def}{=}&
\max\limits_{(x,p)\in U} \sum\limits_{k=1}^{n}
\biggl((f'_{x_k})^2\frac{a}{2b}+(f'_{p_k})^2\frac{b}{2a}\biggr)
\nonumber\\
&=&\max\limits_{(x,p)\in U} \sum\limits_{k=1}^{n}
\left[\biggl(\frac{\partial }{\partial x_k}\biggl( p_j \frac{\partial H}{\partial p_j} \biggr)
\biggr)^2\frac{a}{2b}+
\biggl(\frac{\partial }{\partial p_k}
\biggl( p_j \frac{\partial H}{\partial p_j} \biggr)\biggr)^2\frac{b}{2a}\right].
\nonumber
\end{eqnarray}

Put $C_{2j}\stackrel{def}{=}C'_{2j}+C''_{2j}$. Then using inequalities~(\ref{A_2j_leq}),
(\ref{res'_2j}), and (\ref{res''_2j}), and using the introduced notation
$C_{2j}$, we finally obtain
\begin{eqnarray}\label{res_2j}
 ||A_{2j} \bar \varphi_0\!-\!P_0 A_{2j} \bar \varphi_0||\!\leq\!
\frac{\left(C'_{2j}\!+\!o_{\hbar}(1)\right) }
{\sqrt {\hbar}}\!+\!\frac{\left(C''_{2j}\!+\!o_{\hbar}(1)\right)}{\sqrt {\hbar}}\!=\!
\left(C_{2j}\!+\!o_{\hbar}(1)\right) \frac{1}{\sqrt {\hbar}}.
\end{eqnarray}

This finishes examining Case 3.

Now we are ready to finish the proof of Statement~2.

By Statement~1,
$\alpha_{max}\stackrel{def}{=}
2\cdot  \max_{\bar \varphi_0}|| A \bar \varphi_0- P_0 A\bar\varphi_0||$. Hence,
applying to inequality~(\ref{sum_normA_j}) relations~(\ref{res_2n_1}), (\ref{res_2j_1}),
and (\ref{res_2j}), we obtain
\begin{eqnarray}
\alpha_{max}=2\cdot  \max_{\bar \varphi_0}|| A \bar \varphi_0- P_0 A\bar\varphi_0||\leq
2\cdot  \max_{\bar \varphi_0}
\sum\limits_{s=1}^{2n+1}|| A_s \bar \varphi_0- P_0 A_s\bar\varphi_0||
\nonumber\\
=2 \sum\limits_{s=1}^{2n+1}\left(C_{s}+o_{\hbar}(1)\right) \frac{1}{\sqrt {\hbar}}
=\left(C+o_{\hbar}(1)\right) \frac{1}{\sqrt {\hbar}},\nonumber
\end{eqnarray}
where $C\stackrel{def}{=}2 \sum_{s=1}^{2n+1}C_{s}$.

Q. E. D. Statement 2 is proved.

\newpage
\begin{flushright}
{\large Appendix 2}
\end{flushright}

\section*{Proof of Theorem 5}
\addcontentsline{toc}{section}{Appendix 2. Proof of Theorem 5}

For proof of Theorem~5, consider formula~(\ref{hat_H}) from Theorem~4 for the case when
$H(x, p) = \frac{p^2}{2m}+ V(x).$ We have
\begin{eqnarray}\label{hat_T_V}
\hat{H}\psi  &= & \frac {1}{(2\pi \hbar)^n} \int \limits_{R^{3n}}
\biggl (\frac{p^2}{2m}+ V(x)-\sum_{k=1}^{n}\biggl(\frac{ \partial V}{\partial x_k}
+\frac {i b}{a}\frac{ p_k}{ m }\biggr)(x_k-y'_k)
\biggr) \nonumber\\
   &  & \times \chi(x- y) \chi(x- y')  e^{\frac{i}{\hbar}(y-y') p}  \psi(y',t) dy' dx dp.
\end{eqnarray}

Thus, in this case $\hat{H}\psi $ is represented as the sum of three
integrals
$\hat{H}\psi=I_1+I_2+I_3$, where
\begin{eqnarray}\label{I_1}
I_1& = & \frac {1}{(2\pi \hbar)^n} \int \limits_{R^{3n}}
\frac{p^2}{2m}  \chi(x- y) \chi(x- y')  e^{\frac{i}{\hbar}(y-y') p}  \psi(y',t) dy' dx dp, \\
\label{I_2}
I_2 &=& -\frac {1}{(2\pi \hbar)^n} \int \limits_{R^{3n}}
 \frac {i b}{a m}\sum_{k=1}^{n} p_k(x_k-y'_k) \nonumber\\
   &  & \times \chi(x- y) \chi(x- y')  e^{\frac{i}{\hbar}(y-y') p}  \psi(y',t) dy' dx dp,\\
\label{I_3}
I_3&=& \frac {1}{(2\pi \hbar)^n} \int \limits_{R^{3n}}
\biggr(V(x)-\sum_{k=1}^{n}\frac{ \partial V}{\partial x_k}(x_k-y'_k)\biggl) \nonumber\\
   &  & \times  \chi(x- y) \chi(x- y')  e^{\frac{i}{\hbar}(y-y') p}  \psi(y',t) dy' dx dp.
\end{eqnarray}

Note that the expression $\int_{R^n} \chi(x- y) \chi(x- y') dx$ from the first
integral can be transformed to the following form:
\begin{eqnarray}\label{chi_chi}
\lefteqn{\int\limits_{R^n}\! \chi(x- y) \chi(x- y') dx=
\left(\frac{b}{a\pi\hbar}\right)^{n/2}\!\!\! \int \limits_{R^n}\!
e^{-\frac{{b}(x-y)^2}{2a\hbar}}
e^{-\frac{{b}(x-y')^2}{2a\hbar}}dx}\nonumber\\
&=&\left(\frac{b}{a\pi\hbar}\right)^{n/2}\!\! \!\int \limits_{R^n}\!
e^{-\frac{{b}(x-{(y+y')}/{2})^2}{a\hbar}}
e^{-\frac{{b}(y-y')^2}{4a\hbar}}d\biggl(\!x-\frac{y+y'}{2}\!\biggr)=
e^{-\frac{{b}(y-y')^2}{4a\hbar}}.
\end{eqnarray}

Hence the integral $I_1$  is transformed to the form
\begin{eqnarray}\label{I_1_1}
I_1& = & \frac {1}{(2\pi \hbar)^n} \int \limits_{R^{2n}}
\frac{p^2}{2m}  e^{-\frac{{b}(y-y')^2}{4a\hbar}}
e^{\frac{i}{\hbar}(y-y') p}  \psi(y',t) dy' dp.
\end{eqnarray}

As known from the formulas for Fourier transform, for any smooth function $f(y')$
the following equality holds:
 \begin{equation}\label{f_delta}
f(y)= \frac {1}{(2\pi \hbar)^n} \int \limits_{R^{2n}}
  e^{\frac{i}{\hbar}(y-y') p}  f(y') dy' dp.
\end{equation}
Below we shall often use this equality.

In particular, one has the equality
\begin{equation}
\psi(y,t)=\frac {1}{(2\pi \hbar)^n} \int \limits_{R^{2n}}
  e^{-\frac{{b}(y-y')^2}{4a\hbar}}  e^{\frac{i}{\hbar}(y-y') p}  \psi(y',t) dy' dp.
\end{equation}

Let us differentiate both parts of the latter equality with respect to $y_k$. We obtain,
taking into account relation~(\ref{f_delta}),
\begin{eqnarray}
\frac{\partial \psi(y,t)}{\partial y_k}&
=&\frac {1}{(2\pi \hbar)^n} \int \limits_{R^{2n}}\biggl( \frac{i p_k }{\hbar}
-\frac{{b}(y_k-y'_k)}{2 a\hbar}\biggr)
  e^{-\frac{{b}(y-y')^2}{4a\hbar}}
  e^{\frac{i}{\hbar}(y-y') p}  \psi(y',t) dy' dp\nonumber\\
&=&\frac {1}{(2\pi \hbar)^n} \int \limits_{R^{2n}} \frac{i p_k }{\hbar}
  e^{-\frac{{b}(y-y')^2}{4a\hbar}}  e^{\frac{i}{\hbar}(y-y') p}  \psi(y',t) dy' dp.
\end{eqnarray}
If we differentiate both parts of the obtained equality with respect to $y_k$ once more,
then we obtain the equality
\begin{eqnarray}\label{d2psi}
\frac{\partial^2 \psi(y,t)}{\partial y_k^2}&=&
\frac {1}{(2\pi \hbar)^n} \int \limits_{R^{2n}}
\frac{i p_k }{\hbar}\biggl( \frac{i p_k }{\hbar} -\frac{{b}(y_k-y'_k)}{2 a\hbar}\biggr)
\nonumber\\
 &&\times  e^{-\frac{{b}(y-y')^2}{4a\hbar}}  e^{\frac{i}{\hbar}(y-y') p}  \psi(y',t) dy' dp=
-\frac {2m}{ \hbar^2} I_1^k-I_4^k,
\end{eqnarray}
where
\begin{equation}\label{I_1_k}
I_1^k=\frac {1}{(2\pi \hbar)^n } \int \limits_{R^{2n}}\frac {p_k^2}{2m}
  e^{-\frac{{b}(y-y')^2}{4a\hbar}}  e^{\frac{i}{\hbar}(y-y') p}  \psi(y',t) dy' dp,
\end{equation}
and
\begin{equation}\label{I_4_k}
I_4^k=\frac {1}{(2\pi \hbar)^n} \int \limits_{R^{2n}}\frac{i p_k}{\hbar}
\frac{{b}(y_k-y'_k)}{2 a\hbar}
  e^{-\frac{{b}(y-y')^2}{4a\hbar}}  e^{\frac{i}{\hbar}(y-y') p}  \psi(y',t) dy' dp.
\end{equation}

To compute the integrals $I_4^k$, consider the following equality which is a particular
case of equality~(\ref{f_delta}):
\begin{equation}
0=\frac {1}{(2\pi \hbar)^n} \int \limits_{R^{2n}}  \frac{{b}(y_k-y'_k)}{2 a\hbar}
  e^{-\frac{{b}(y-y')^2}{4a\hbar}}  e^{\frac{i}{\hbar}(y-y') p}  \psi(y',t) dy' dp.
\end{equation}
Let us differentiate both parts of this equality with respect to $y_k$. We obtain
\begin{eqnarray}
0&=&\frac {1}{(2\pi \hbar)^n} \int \limits_{R^{2n}}
 \biggl(\frac{b}{2 a\hbar}+ \frac{{b}(y_k-y'_k)}{2 a\hbar}
 \biggl( \frac{i p_k }{\hbar} -\frac{{b}(y_k-y'_k)}{2 a\hbar}\biggr)\biggr)
\nonumber\\
&&\times e^{-\frac{{b}(y-y')^2}{4a\hbar}}  e^{\frac{i}{\hbar}(y-y') p}  \psi(y',t) dy' dp.
\end{eqnarray}
Hence, taking into account notation~(\ref{I_4_k}) for the integral $I_4^k$
and relation~(\ref{f_delta}), we obtain the equalities
\begin{equation}\label{I_4_res}
0=\frac{b}{2 a\hbar}\psi(y,t)+I_4^k \ \ \ \mbox{ or }\ \ \  I_4^k=-\frac{b}{2 a\hbar}\psi(y,t).
\end{equation}

Substituting this equality into relation~(\ref{d2psi}), we obtain
$$\frac{\partial ^2 \psi}{\partial y_k^2}
=-\frac {2m}{\hbar^2} I_1^k+\frac{b}{2 a\hbar}\psi(y,t).$$
Let us express $I_1^k$ from this equality:
$$
I_1^k= -\frac{\hbar^2}{2m}\frac{\partial ^2 \psi}{\partial y_k^2}+\frac{b\hbar}{4m a}\psi(y,t).
$$
Taking the sum of the obtained equality over all $k$ from 1 to $n$, we obtain an expression
for the integral $I_1$:
\begin{equation}\label{I_1_res}
I_1=\sum_{k=1}^n I_1^k=-\sum_{k=1}^n
\frac{\hbar^2}{2m}\frac{\partial ^2 \psi}{\partial y_k^2}+\frac{nb\hbar}{4m a}\psi(y,t).
\end{equation}

Let us pass to computing the integral $I_2$ given by expression~(\ref{I_2}).

Consider equality~(\ref{chi_chi}) of the form
$$
\int\limits_{R^n}\! \chi(x- y) \chi(x- y') dx \!=\!
 e^{-\frac{{b}(y-y')^2}{4a\hbar}},\ \ \
\mbox{where}\ \ \
\chi(x-y')\!=\!\left(\frac{b}{a\pi\hbar}\right)^{n/4}\!e^{-\frac{{b}(x-y')^2}{(2a\hbar)}}.
$$
Let us differentiate both parts of this equality with respect to $y'_k$. We obtain
$$
\int\limits_{R^n}\!\frac{b(x_k-y'_k)}{a\hbar} \chi(x- y) \chi(x- y') dx= \frac{{b}(y_k-y'_k)}{2a\hbar}
 e^{-\frac{{b}(y-y')^2}{4a\hbar}},
$$
or, after omitting common factors,
\begin{equation}
\int\limits_{R^n}(x_k-y'_k) \chi(x- y) \chi(x- y') dx= \frac{y_k-y'_k}{2}
 e^{-\frac{{b}(y-y')^2}{4a\hbar}}.\nonumber
\end{equation}

Substituting this equality into expression~(\ref{I_2}) for the integral $I_2$, we obtain
$$
I_2 = -\frac {1}{(2\pi \hbar)^n} \int \limits_{R^{2n}}
 \frac {i b}{2a m}\sum_{k=1}^{n} p_k(y_k-y'_k)
e^{-\frac{{b}(y-y')^2}{4a\hbar}} e^{\frac{i}{\hbar}(y-y') p}  \psi(y',t) dy' dp.
$$
Comparing the latter expression with expression~(\ref{I_4_k}) for the integrals $I_4^k$,
we obtain
$I_2 = - {\hbar^2}/{ m}\sum_{k=1}^{n}I_4^k .$
Substituting here the computed expressions~(\ref{I_4_res}) for the integrals $I_4^k$,
we finally obtain
\begin{equation}\label{I_2_res}
I_2 = \frac {n b\hbar}{2 ma}\psi(y,t).
\end{equation}

Now consider the integral $I_3$ given by expression~(\ref{I_3}).
In accordance with formula~(\ref{f_delta}),  this integral can be transformed to the form
$$
I_3=\psi(y,t) \int \limits_{R^{n}}
\biggr(V(x)-\sum_{k=1}^{n}\frac{ \partial V}{\partial x_k}(x_k-y_k)\biggl)
     \chi^2(x-y)     dx ,$$
where $\chi^2(x-y)=({b}/(a\pi\hbar))^{n/2}exp({-{{b}(x-y)^2}/{(a\hbar)}})$
is the density of the normal distribution, or, after the change of variables
$x'_k=x_k-y_k$, to the form
$$
I_3=\psi(y,t) \int \limits_{R^{n}}
\biggr(V(y+x')-\sum_{k=1}^{n}\frac{ \partial V(y+x')}{\partial y_k} x_k'\biggl)
     \chi^2(x')     dx'. $$

Whereas the previous integrals have been computed exactly, let us compute this integral
approximately, assuming the dispersion ${a\hbar}/{2b}$ of the normal distribution
$\chi^2(x')$ to be a small quantity,  decomposing the function $V(y+x')$ into the Taylor
series at the point $y$ with respect to $x'$ up to the second order, and decomposing
     $ {\partial V(y+x')}/{\partial y_k}$ up to the first order. We have
\begin{eqnarray}
 I_3\approx \psi (y,t)
\int\limits_{R^n}\left( V(y)+
\sum_{k=1}^n\frac{\partial V(y)}{\partial y_k}x'_k+
\frac{1}{2}\sum_{k=1}^n\sum_{k'=1}^n
\frac{\partial^2 V(y)}{\partial y_k\partial y_{k'}}x'_{k}x'_{k'}
\right)\chi^2(x')dx'&&
\nonumber\\
-\psi(y,t)\int\limits_{R^n}\left(\sum_{k=1}^n
\frac{\partial V(y)}{\partial y_k}x'_k+
\sum_{k=1}^n\sum_{k'=1}^n \frac{\partial^2 V(y)}{\partial y_k\partial y_{k'}}x'_{k}x'_{k'}
\right)\chi^2(x')dx'.&&
\nonumber
\end{eqnarray}
Hence, since for the density of the normal distribution $\chi^2(x')$ the following relations
hold: $\int_{R^n}\chi^2(x') dx'=1,$
$\int_{R^n}x'_k\chi^2(x')dx'=0,$  $\int_{R^n}x'_k x'_{k'}\chi^2(x') dx'=0$ for $k\not =k'$, and
$\int_{R^n}x'_k x'_k\chi^2(x') dx'=a\hbar/(2b)$,  we finally obtain
\begin{eqnarray}\label{I_3_res}
\ I_3\approx
\psi(y,t)V(y)-\psi(y,t)\frac{a\hbar}{4b}
\sum_{k=1}^n
\frac{\partial^2 V(y)}{\partial y_k^2}.
\end{eqnarray}

Thus, since $\hat H\psi = I_1+I_2+I_3$, where the integrals $I_1, I_2, I_3$ are given
respectively by expressions
(\ref{I_1}), (\ref{I_2}), (\ref{I_3}), then, substituting here their computed
values as expressions (\ref{I_1_res}), (\ref{I_2_res}), (\ref{I_3_res}) and reducing
similar summands, we obtain the expression for $\hat H$ required in Theorem~5:
\begin{equation}
\hat H \approx - \frac{\hbar^2}{2m}\biggl(\sum_{k=1}^{n}\frac{\partial^2 }
{\partial{y^2_k}}\biggr)
+V(y)-\frac{a\hbar}{4b}\sum_{k=1}^{n}
\frac{\partial^2 V}{\partial{y^2_k}} +\frac{3nb\hbar}{4ma}.\nonumber
\end{equation}

\newpage
\begin{flushright}
{\large Appendix 3}
\end{flushright}

\section*{An estimate of parameters of the model}
\addcontentsline{toc}{section}{Appendix 3. An estimate of parameters of the model}

\section*{1. An estimate of the parameter $a/b$ of the model}

The difference between the operator $\hat H$ given by expression~(\ref{hatH}) of Theorem~5
and the standard Hamilton operator in the Schrodinger equation, can cause difference between
the spectra of the energy operators (in particular, for the potential energy function
of hydrogen atom) for the model considered above and
for the standard model of quantum mechanics.
The difference between these operators is in the third summand with the factor
${a\hbar}/{4b}$. If one assumes that behavior of particles is described by
the Schrodinger equation with the operator $\hat H$ of the form~(\ref{hatH}) more exactly
than with he Hamilton operator, then experiments should show the non-exactness of spectra
computed by means of the Hamilton operator, i.~e., non-exactness
of non-relativistic quantum mechanics.

The discrepancy between theoretical and experimental data in non-rel\-at\-iv\-ist\-ic
quantum mechanics is well known. It has been discovered in 40s in
\cite{lamb}, and has been called the Lamb shift of levels of hydrogen atom.
Later on, this effect has been explained in quantum electrodynamics by interaction
of the electron with fluctuating electromagnetic field (see,  for example, \cite{landau4},
p.~593, where one can find references to original works).

If one assumes that the data in the Lamb experiment are related with the perturbing
summand in the operator $\hat H$,
then these data allow one to estimate the quantity ${a\hbar}/{4b}$.

In this section the computations of the estimate of the quantity ${a\hbar}/{4b}$
have been carried out following the computations given in~\cite{welt} for
substantiation of the size of the Lamb shift in the spectrum of hydrogen atom.

Let
$$V(y)=-\frac{e^2}{r}=-\frac{e^2}{\sqrt{y_1^2+y_2^2+y_3^2}} $$
be the potential function of hydrogen atom, and $ \hat H = \hat E + \hat V$ be the
operator from Theorem~5, where
$$
\hat E= - \frac{\hbar^2}{2m}\biggl(\sum_{k=1}^{3}\frac{\partial^2 }{\partial{y^2_k}}\biggr)
\ \ \ \mbox {and}\ \ \
\hat V(y)=V(y)-\frac{a\hbar}{4b}\sum_{k=1}^{3}\frac{\partial^2 V}{\partial{y^2_k}}.
$$

Consider the operator $\hat H$  as a perturbation of the Hamilton operator
$ \tilde H =  \hat E +V$ of hydrogen atom. Let us estimate eigenvalues of the operator
$\hat H. $

The standard perturbation theory implies that at the first approximation, the
correction $\delta E_n$
to the eigenvalue $E_n$   of the Hamilton operator has the form
$$\delta E_n= \int_{R^3}{\rho_n(y) (\hat V(y)-V(y) )dy},$$
where $\rho_n(y)=|\psi_n(y)|^2$ and $\psi_n(y)$ is the eigenfunction of the Hamilton
operator with the eigenvalue $E_n$.

Substituting the expression for $\hat V(y)$ into the expression for $\delta E_n$, we obtain
$$\delta E_n= - \frac{a\hbar}{4b}\int_{R^3}{\rho_n(y) \sum_{k=1}^3
\frac{\partial^2 V(y)}{\partial y_k^2}dy}.$$

Since the integral of the Laplace operator of the function $V(y)=-{e^2}/{r}$
equals $4 \pi e^2 {\delta}_0{(y)},$   where $4 \pi e^2 {\delta}_0{(y)}$ is the delta function
at zero, this implies that
$$ \delta E_n= - \frac{a\hbar \pi e^2}{b} \rho_n(0). $$

For hydrogen atom it is known (see, for example,~\cite{sok}, p.~342) that
 $$\rho_n(0)=|\psi_n(0)|^2=\frac{1}{\pi n^3}\left(\frac{m e^2}{\hbar^2} \right)^3.$$
Hence
$$ \delta E_n
= - \frac{a\hbar \pi e^2}{b} \frac{1}{\pi n^3}\left(\frac{m e^2}{\hbar^2} \right)^3=
-\frac{a}{b}\frac{m^3 e^8}{n^3 \hbar^5}=
-\frac{a}{b}\frac{m^3 \alpha^4 c^4}{n^3 \hbar},$$
where $\alpha={e^2}/(\hbar c)={1}/{137}$ is the fine structure constant,  $c$  is the
velocity of light. Hence, $a/b$ is expressed as follows:
$$\frac{a}{b}=|\delta E_n|\frac{n^3\hbar}{m^3 \alpha^4 c^4}.$$
In the Lamb--Retherford experiments for hydrogen atom it has been
established that $\delta E_2   = 1058 MHz = 1058 \cdot 10^6  h $ erg, where $h=2\pi \hbar$.
Comparing this value with the obtained value of $\delta  E_2,$ we obtain by
simple calculations the estimate of the quantity $a/b  = 3,41\cdot 10^4 sec/g.$
Hence the standard deviation for the density of normal probability distribution
$\chi_1^2(x'),$ with which the smoothing of the potential
$V$ is made, equals $\sqrt{a\hbar/2b}=4,24\cdot 10^{-12}cm$. This quantity is much less
than the radius of hydrogen atom.

\section*{2. An estimate of the diffusion coefficients
and of the time of the transformation process}
In this section we shall assume that the diffusion coefficients
$a$ and $b$ are defined by the standard heat action of the surrounding medium on the
moving electron.
One can assume that the Brownian particle (the electron) is acted on by a fluctuating
force from the surrounding medium, and also by stochastic resistance proportional to the
velocity of the particle. For modelling of the motion in this situation, one usually uses
(see, for example, \cite{isihara}, p.~196) the Langevene equation
$$\dot p=-\gamma p +F(t),$$
where $\gamma$ is the friction coefficient for the unit mass, and $F(t)$ represents the
fluctuating force, which is assumed to be independent of the velocity,
with the mean value equal to zero.

This equation is equivalent (see \cite{isihara}, p.~212) to the Fokker--Planck equation
for the density of probability distribution $f(p,q,t)$ in the phase space, in its
standard form:
$$\frac{\partial f}{\partial t}+\frac{p_i}{m}\frac{\partial f}{\partial x_i}=
\gamma \frac{\partial (fp_i)}{\partial p_i}+\gamma kTm\frac{\partial^2 f}{ \partial p_i^2},$$
where $m$ is the mass of the particle (in the case of electron, $m= 9.10939\cdot 10^{-31}kg$),
$k$ is the Boltzmann constant ($k= 1.38066\cdot 10^{-23}J/K$), $T$
is the temperature of the medium.

The solution of these equations for time intervals $t$ much greater than $\gamma^{-1}$,
is well known (see \cite{van}, p.~215) to yield the diffusion process with respect to
coordinates $x$ with the diffusion coefficient $kT/(m\gamma)$. This relation is also known
as the Einstein relation for the diffusion coefficient (\cite{isihara}, p.~198).

Thus, under these assumptions one can suppose that the diffusion coefficient with
respect to coordinates $a^2= kT/(m\gamma)$, and the diffusion coefficient with respect
to momenta (as in the Fokker--Planck equation)
$b^2= \gamma kTm$.

Hence, $a/b=(\gamma m)^{-1}$ and $ab=kT$.

By the estimate obtained in the previous section, we have
 $$\frac{a}{b}=\frac {1}{\gamma m}= 3.41\cdot 10^7 {sec}/{kg},$$ and the quantity
$$\gamma= \frac{1}{3.41\cdot 10^7 \cdot m }=
\frac{1}{3.41\cdot 10^7 \cdot 9.10939\cdot 10^{-31}}=3.22 \cdot 10^{22} sec^{-1}.$$

On the other hand, the time of the transformation process to the
process described by the Schrodinger equation,
is expressed, by Theorem~4, by the quantity ${\hbar}/({ab})={\hbar}/({kT}).$
This time for $T=1^\circ K$ equals $7.638\cdot 10^{-12} sec.$

For the same temperature, the diffusion coefficients are estimated in the following way:
$$a^2= kT/(m\gamma)=4.708 \cdot 10^{-16} m^2\cdot sec^{-1};$$
$$  b^2= \gamma kTm=4.049\cdot 10^{-31}(kg\cdot m\cdot sec^{-1})^2\cdot sec^{-1}.$$


\newpage
\begin{thebibliography}{99}
\addcontentsline{toc}{section}{References}
\bibitem{ben}
{\it Beniaminov E. M.}
{\it Diffusion Processes in Phase Spaces and Quantum Mechanics}
 // Doklady Mathematics. 2007. V. 76,  ¹2, P. 771-774. arXiv:0803.2669v1[math-ph].
\bibitem{ben_conf}
{\it Beniaminov E. M.}
{\it Quantization as approximate description of a diffusion process} // in: Proceedings
of the international workshop ``Idempotent and Tropical Mathematics and Problems of
Mathematical Physics'', Moscow, August 2007, vol.2, p.~78--84 (in Russian).
\bibitem{wigner}
{\it Wigner~E.}
{\it On the Quantum Correction For Thermodynamic Equilibrium}
 // Phys. Rev. 1932. V. 40. P. 749--759.
\bibitem{bohm_vigier}
{\it Bohm~D., Vigier~ J.P.}
{\it  Model of the Causal Interpretation of Quantum Theory in Terms of a Fluid with
Irregular Fluctuations}
// Phys. Rev. 1954. V. 96. P. 208--216.
\bibitem{nelson}
{\it Nelson~E.}
{\it Derivation of the Schrodinger Equation from Newtonian Mechanics,}
// Phys. Rev. 1966. V. 150. P. 1079--1085.
\bibitem{pena_cetto}
{\it De la Pena-Auerbach~L., Cetto~A.M.}
{\it Derivation of quantum mechanics from stochastic electrodynamics}
// J. Math. Phys. 1977. V. 18. P. 1612--1622.
\bibitem{baublitz}
{\it Baublitz~M.}
{\it Derivation of the Schrodinger Equation from a Stochastic Theory}
// Prog. Theor. Phys. V. 80. P. 232--244.
\bibitem{maslov1}
{\it Maslov~V. P.}
{Kolmogorov--Feller equations and a probabilistic model of quantum mechanics}
 // Itogi nauki i tehniki. Probability theory, mathematical statistics, and cybernetics.
 1982. V. 19. P. 55--85 (in Russian).
\bibitem{maslov2}
{\it Maslov V. P.} {Quantization of thermodynamics and ultrasecondary quantization.}
 Moscow, Institute for Computer Studies, 2001. 384 pp. (in Russian)
\bibitem{lamb}
{\it Lamb~W. E., Retherford~R. C.}
{\it Fine Structure of the Hydrogen Atom by a Microwave Method}
 // Phys. Rev. 1947. V. 72. P. 241--243.
\bibitem{feinman}
{\it Feynman R. P., Hibbs A. R.},
{Quantum Mechanics and Path Integrals}, McGraw-Hill, New York, 1965; Mir, Moscow, 1968.
\bibitem{maslov_mark}
{\it Maslov V. P.} {Complex Markov chains and Feynman path integral.}
 Nauka, Moscow, 1976. 192 pp. (in Russian)
\bibitem{landau3}
{\it Landau L. D., Lifschitz E. M.}, {Quantum mechanics (non-relativistic theory).}
Theoretical physics, vol. 3, Nauka, Moscow, 1989 (in Russian).
\bibitem{math}
{Wolfram Mathematica (system for symbolic mathematical computations).} http://www.wolfram.com/
\bibitem{ben_arxiv}
{\it Beniaminov~E. M.}
{\it A Method for  Justification of the View of Observables in Quantum Mechanics and Probability
Distributions in Phase Space.} 2001. arXiv:quant-ph/0106112v1.

\bibitem{landau4}
{\it Berestetsky V. B.,  Lifschitz E. M.,  Pitaevsky~L.~P.} {Quantum
electrodynamics.} Theoretical physics. Vol.~4. Pergamon, Oxford, 1982;
Nauka, Moscow, 1980.
\bibitem{isihara}
{\it Isihara A.}, {Statistical Physics}, Academic, New York, 1971; Mir, Moscow, 1973.
\bibitem{welt}
{\it Welton~T.A.}
 {\it Some Observable Effects of The Quantum-Mechanical Fluctuations of the
 Electromagnetic Field}
 // Phys. Rev. 1948. V. 74. P. 1157--1167.
\bibitem{dyihne}
{\it Dykhne A. M., Yudin G. L.}
{Sudden perturbations and quantum evolution.} Editorial board of
 ``Uspekhi fizicheskih nauk'' (``Russian Phys. Surveys''),
 Moscow, 1996. 428 pp. (in Russian)
\bibitem{sok}
{\it Sokolov A. A., Ternov I. M., Zhukovsky~V.~Ch.} {Quantum mechanics.} Nauka, Moscow, 1979
(in Russian).
\bibitem{van}
{\it Van Kampen N. G.} {Stochastic processes in physics and chemistry.}
North-Holland, Amsterdam, 1984.

\end{thebibliography}
\end{document}